%% file: main_arkiv.tex
\documentclass[sigconf]{acmart}
\acmConference[ESEC/FSE 2022]{The 30th ACM Joint European Software Engineering Conference and Symposium on the Foundations of Software Engineering}{14 - 18 November, 2022}{Singapore}
\usepackage{enumitem}
\usepackage{graphicx}
\usepackage{subfigure}
\usepackage{multirow}
\usepackage{hhline}
\usepackage{colortbl}
\usepackage{multicol}
\usepackage{makecell}
\usepackage{fancyhdr}
\usepackage{xcolor}
\usepackage{array}
\usepackage{soul}
\usepackage{listings}
\usepackage[normalem]{ulem}
\usepackage{algorithm,algcompatible,amsmath}
\makeatletter

\newtoks\therules
\therules={}
\def\appendto#1#2{\expandafter#1\expandafter{\the#1#2}}
\def\gobblefirst#1{
  #1\expandafter\expandafter\expandafter{\expandafter\@gobble\the#1}}%
\def\LState{\State\unskip\the\therules}
\def\printindent{\unskip\the\therules}%
\renewcommand\footnotetextcopyrightpermission[1]{} 
\AtBeginDocument{%
  \providecommand\BibTeX{{%
    \normalfont B\kern-0.5em{\scshape i\kern-0.25em b}\kern-0.8em\TeX}}}

\usepackage{framed}  
\usepackage{lipsum}
\usepackage{xcolor}
\usepackage{arydshln} 
\renewcommand\fbox{\fcolorbox{blue}{white}}
\usepackage{hyperref}

\let\underscore\_
\renewcommand{\_}{\discretionary{\underscore}{}{\underscore}}

\newcommand{\zc}[1]{\textcolor{red}{\textbf{ZC:} #1}}
\newcommand{\zj}[1]{\textcolor{blue}{\textbf{Zejun:} #1}}
\begin{document}

\title{Making Python Code Idiomatic by Automatic Refactoring Non-Idiomatic Python Code with Pythonic Idioms
}

\author{Zejun Zhang}
\affiliation{%
  \institution{Australian National University}
  \country{Australia}}
\email{zejun.zhang@anu.edu.au}
\author{Zhenchang Xing}
\affiliation{%
  \institution{Data61, CSIRO}
  \institution{Australian National University}
  \country{Australia}}
\email{zhenchang.xing@anu.edu.au}

\author{Xin Xia}\authornote{Corresponding author.}
\affiliation{%
  \institution{Software Engineering Application Technology Lab, Huawei}
  \country{China}
}
\email{xin.xia@acm.org}

\author{Xiwei Xu}
\affiliation{%
  \institution{Data61, CSIRO}
  \country{Australia}}
\email{xiwei.xu@data61.csiro.au}
\author{Liming Zhu}
\affiliation{%
  \institution{Data61, CSIRO}
  \country{Australia}}
\email{Liming.Zhu@data61.csiro.au}





\begin{abstract}
Compared to other programming languages (e.g., Java), Python has more idioms to make Python code concise and efficient.
Although pythonic idioms are well accepted in the Python community, Python programmers are often faced with many challenges in using them, for example, being unaware of certain pythonic idioms or do not know how to use them properly. 
Based on an analysis of 7,638 Python repositories on GitHub, we find that non-idiomatic Python code that can be implemented with pythonic idioms occurs frequently and widely. 
Unfortunately, there is no tool for automatically refactoring such non-idiomatic code into idiomatic code. 
In this paper, we design and implement an automatic refactoring tool to make Python code idiomatic.
We identify nine pythonic idioms by systematically contrasting the abstract syntax grammar of Python and Java. 
Then we define the syntactic patterns for detecting non-idiomatic code for each pythonic idiom. 
Finally, we devise atomic AST-rewriting operations and refactoring steps to refactor non-idiomatic code into idiomatic code. 
We test and review over 4,115 refactorings applied to 1,065 Python projects from GitHub, and submit 90 pull requests for the 90 randomly sampled refactorings to 84 projects.
These evaluations confirm the high-accuracy, practicality and usefulness of our refactoring tool on real-world Python code. 
Our refactoring tool can be accessed at \href{47.242.131.128:5000}{\textcolor{blue}{47.242.131.128:5000}}.

\end{abstract}



\keywords{Pythonic Idioms, Abstract Syntax Grammar, Code Refactoring}


\maketitle
\section{Introduction}
\input{intro_3}

\section{Formative Study and Motivation}\label{motivation}
\input{motiv_3_arkiv.tex}

\section{Our Approach}\label{method}
\input{method_3_arkiv}

\section{Evaluation}\label{result}
\input{result_3_arkiv}

\section{Related Work}\label{relatedwork}
\input{related}

\section{Discussion}\label{threats}

\subsection{Pythonic Coding Practices}
Refactoring is a widely adopted practice to improve code quality.
A wide range of refactorings have been proposed to address code smells such as code clones, feature envy, shotgun surgery.
Our work introduces a new type of code smell, i.e., non-idiomatic code that can be refactored with pythonic idioms.
Our empirical study on GitHub repositories and Stack Overflow questions calls for the tool support for assisting developers in using pythonic idioms consistently.
Our refactoring tool is the first tool of this kind. 
The evaluation on a large number of Python projects provides positive and encouraging feedback on the prototype.
Some developer feedback raises concerns about the readability and performance of pythonic idioms.
This calls for the careful validation of the conciseness and performance of pythonic idioms.
However, existing online materials are anecdotal and mostly based on personal programming experience.
Our work produces a large dataset of non-idiomatic versus idiomatic code from real-world projects, which serves as an excellent testbed to empirically investigate the general claims and concerns about pythonic idioms.

\subsection{Threats to Validity}

\textbf{Threats to internal validity} relate to two aspects in our work: (1) the errors in the implementation of code refactoring tool and (2) personal bias in evaluating accuracy of code refactoring.
For the aspect (1), we have double-checked the code and verified the accuracy of our tool implementation by manually examining a large number of refactoring instances outputted by each step of our tool.
As for the aspect (2), two authors with more than three years of Java and Python programming experience check the accuracy of refactoring instances independently. Furthermore, we collect a large number of real test cases to test the refactored code. 
\textbf{Threats to external validity} relate to the generalizability of experiment results. To alleviate this threat, we built a large-scale dataset of 7,638 repositories and 506,765 Python files. To explore whether our code refactoring has practical value for developers, we submitted 90 pull requests to project members to review. The number of pull requests is larger than existing user studies in previous works~\cite{gao2021automating,pan2020reinforcement,zhang2020chatbot4qr}. 
We release our tool and data in Zenodo\footnote{https://zenodo.org/record/6367738\#.YjRzLxBBzdo} for public evaluation.

\section{Conclusion and Future Work}\label{conclusion}
This paper designs and implements the first automatic refactoring tool for nine types of pythonic idioms.
Our tool is motivated by the empirical observation of the challenges in writing pythonic code from the Stack Overflow discussions and of the wide presence of non-idiomatic code in thousands of real-world Python projects.
Rather than relying on idiom mining, literature review or personal programming experience, our approach identifies pythonic idioms and define  non-idiomatic syntactic patterns and idiomatic code transformation steps through the systematic analysis of Python abstract syntax grammar.
Our tool is robust and correct in detecting anti-idiom code smells and refactor these smells in real-world Python projects.
The refactorings made by our tool have been well received and praised by the Python developers.
In the future, we will integrate our refactoring tool into the open-source linting tool (e.g., Pylint).
We will systematically investigate the readability and performance concerns about pythonic idioms based on the large-scale refactoring dataset our tool produces.

\balance
\bibliographystyle{ACM-Reference-Format}
\bibliography{sample-base}




\end{document}

%% file: intro_3.tex

Programming (or code) idioms are widely present in programming languages~\cite{programming_idioms}.
They represent notable programming styles and features of a programming language.
Python is well known for its pythonic idioms~\cite{van2007python,merchantepython}.
Many books and online materials~\cite{alexandru2018usage,reitz2016hitchhiker,knupp2013writing,hettinger2013transforming,slatkin2019effective,merchantepython} promote the use of pythonic idioms for not only concise coding styles but also improved performance (see Table~\ref{tab:benefits} for examples).
In spite of the benefits of pythonic idioms and the availability of many online materials, our investigation of some highly viewed Python questions on Stack Overflow suggests that developers are often unaware of pythonic idioms or do not know when and how to use pythonic idioms properly (see the examples in Table~\ref{tab:motivation_soposts} and the analysis in Section~\ref{sec:challenges}).


Due to these challenges in using pythonic idioms, developers may implement a functionality in a non-idiomatic way without using pythonic idioms.
Table~\ref{tab:benefits} shows some examples.
We study 7,638 Python projects on GitHub (see Section~\ref{sec:challenges}) and find that non-idiomatic code that can be implemented with pythonic idioms is widely present at the repository, file, method and statement level (see Table~\ref{tab:codingpractices}).
Non-idiomatic code and idiomatic code co-exist in many repositories, files or even methods.
As seen in Table~\ref{tab:benefits}, non-idiomatic code syntax is similar to those of other mainstream programming languages (e.g., Java).
In contrast, pythonic idioms have ``uncommon'' syntax.
Developers, even those with little Python programming experience, can still write non-idiomatic code.
However, to use pythonic idioms, they would need to learn new syntax or need some tool supports.

Although online documentation of pythonic idioms provide rich learning materials, they cannot directly support programming with pythonic idioms.
To the best of our knowledge, only two tools provide limited support for using pythonic idioms. 
Among the nine pythonic idioms in Table~\ref{tab:benefits},
Pylint~\cite{Pylint} can detect two types of non-idiomatic code which can be refactored into chain-comparison and truth-value-test respectively.
However, it offers only a simple refactoring suggestion which may not be intuitive to developers.
For example, for the code \textsf{``a < 0 and b > 1 and c < 1 and d > 2''}, Pylint suggests ``Simplify chained comparison between the operands''\footnote{\href{https://github.com/PyCQA/pylint/issues/5800}{https://github.com/PyCQA/pylint/issues/5800}}. 
Unfortunately, the developer did not understand this suggestion initially.
When Pylint developer further explained that the code can be refactored into \textsf{``a < 0 and b > 1 > c and d > 2''}, the developer understood what to do and left a comment ``Would it be an idea for Pylint to output the suggested refactor to the user?'' which received thumbs up by other developers.
Teddy~\cite{phan2020teddy} collects 58 non-idiomatic code fragments and corresponding 55 idiomatic code (three pythonic idioms list-comprehension, set-comprehension and truth-value-test overlap with 9 idioms in Table~\ref{tab:benefits}).
It detects non-idiomatic code similar to the collected 58 examples and recommends corresponding idiomatic code examples.
Developers still have to manually refactor the non-idiomatic code.

In this paper, we develop the first automatic refactoring tool that detects 9 types of non-idiomatic code (referred to as anti-idiom code smells) and refactors these anti-idiom code smells into idiomatic code implementing the same functionalities. 
Existing work~\cite{merchantepython,sivaraman2021mining,allamanis2018mining,allamanis2014mining,Zen_Your_Python} mines recurring code idioms from source code or relies on books that include many idioms that are not unique to Python.
To identify unique pythonic idioms, we contrast the language syntax of Python and the other mainstream programming language (Java in this work).
This is inspired by the observation that non-idiomatic code syntax is similar to those of other languages but idiomatic code has unique syntax.
Our analysis identifies 9 types of pythonic idioms (see Table~\ref{tab:benefits}).
We confirm the validity of these pythonic idioms through the Python language specification and online materials~\cite{knupp2013writing,hettinger2013transforming,slatkin2019effective}.
For each pythonic idiom, we define syntactic patterns for detecting non-idiomatic code fragments that implement the same functionality as the pythonic idiom.
Following the refactoring principle (one small step at a time)~\cite{fowler2018refactoring}, we formulate four atomic AST rewriting operations and compose these atomic operations for each pythonic idiom for refactoring anti-idiom code with the corresponding pythonic idiom.

To evaluate the code smell detection and refactoring accuracy of our approach, we apply our tool to 7,638 Python projects which detects and refactors over 2,252,022 anti-idiom code smells.
We verify the refactoring results by both testing and code review.
Our approach achieves 100\% smell detection accuracy for six idioms and 100\% refactoring accuracy for eight idioms.
It makes only a few rare detection errors and only one refactoring error due to the limitation of Python static analysis and the complex program logic.
To explore the usefulness of our code refactoring tool in practice, we randomly sample 10 refactorings respectively for each pythonic idiom, and submit in total 90 pull requests to 84 projects to make the project members review our refactorings. 
As a result, we receive 57 replies from 54 projects, of which 34 accept our pull request with praise of our refactorings and 28 replies merge the pull requests into their repositories. 
Our results show developers care about pythonic idiom refactorings, and our refactorings have been well received in practice.
The developers' feedback on the rejected pull requests reveal some interesting concerns about the readability and performance of pythonic idioms which deserve further study.
The dataset of anti-idiom code and corresponding idiomatic code produced in this work provides the first large-scale test bed to systematically investigate such concerns.

In summary, this paper makes the following contributions:

\begin{itemize}[leftmargin=*]
  \item To the best of our knowledge, we are the first to automatically detect non-idiomatic code and refactor it into idiomatic code for 9 widely used pythonic idioms.
  
  \item Through the evaluation on a large number of real-world Python projects, we confirm the high accuracy, practicality and usefulness of our refactoring tool. 
  
  \item Our work creates the first large-scale dataset of anti-idiom code smells and corresponding idiomatic code for studying and validating the claims and concerns about pythonic idioms.

\end{itemize}

%% file: motiv_3_arkiv.tex
We conduct an empirical study of pythonic coding practices to answer the following three research questions: 
\begin{table}[htbp]
\setlength\tabcolsep{3.6pt}%
\scriptsize\caption{Pythonic idioms: conciseness and performance}

\vspace{-0.4cm}
  \label{tab:benefits}
\centering
\begin{tabular}{|l|l|l|l|c|} 
\hline
Idiom          & Resource & Anti-idiom code smell &Idiomatic code&X                                                                                                             \\ 
\hline
\begin{tabular}[c]{@{}l@{}}List \\Compre-\\hension\end{tabular}       & \begin{tabular}[c]{@{}l@{}}performance benefits to \\using a list compre-\\hension~\cite{knupp2013writing,list_comprehension_performance,alexandru2018usage}\end{tabular} & 
\begin{tabular}[c]{@{}l@{}}  
t = []\\
for i in range(10000):\\
\ \     t.append(i)
\end{tabular}             & 
\begin{tabular}[c]{@{}l@{}}
t =[i for i in \\range(10000)]
\end{tabular}&2.07  \\ 
\hline
\begin{tabular}[c]{@{}l@{}}Set \\Compre-\\hension\end{tabular}        & \begin{tabular}[c]{@{}l@{}}Set comprehension \\is a more compact \\and faster way to \\create sets~\cite{set_comprehension_performance}\end{tabular}& 
\begin{tabular}[c]{@{}l@{}}
simpsons\_set =set()\\
for word in chars:\\
\ \ simpsons\_set.add(word)

\end{tabular}             & 
\begin{tabular}[c]{@{}l@{}}
simpsons\_set = \{\\ word for word \\in chars\}
\end{tabular} &
1.70           \\ 
\hline
\begin{tabular}[c]{@{}l@{}}Dict \\Compre-\\hension\end{tabular}       & \begin{tabular}[c]{@{}l@{}}Use a dict comp-\\rehension to build \\a dict clearly and\\ efficiently~\cite{knupp2013writing,alexandru2018usage,dict_comprehension_performance}\end{tabular} & \begin{tabular}[c]{@{}l@{}}
b = \{\}\\
for k,v in a.items():\\
\ \     b[v]=k \end{tabular}           & 
\begin{tabular}[c]{@{}l@{}}
b = \{v: k \\for k, v in a.items()\}
\end{tabular}    &1.09        \\ 
\hline
\begin{tabular}[c]{@{}l@{}}Chain \\Compar-\\ison\end{tabular}         & \begin{tabular}[c]{@{}l@{}}...can have a posi-\\tive effect on per-\\formance~\cite{knupp2013writing,chainCompare_perfor_wiki,chainCompare_so}\end{tabular} & 
\begin{tabular}[c]{@{}l@{}}
a <= b \\and b <= c and c <= d \\and d <= e and e <= f
\end{tabular}             & 
\begin{tabular}[c]{@{}l@{}}
a <= b <= c \\<= d <= e <= f
\end{tabular} &1.15    \\ 
\hline
\begin{tabular}[c]{@{}l@{}}Truth \\Value\\ Test \end{tabular}       & \begin{tabular}[c]{@{}l@{}}...can make your \\code more effici-\\ent...~\cite{truth_so, list_truth_so,writing_fast_python}\end{tabular} & \begin{tabular}[c]{@{}l@{}}
if a==[]:\\
\ \ pass\end{tabular}             &
\begin{tabular}[c]{@{}l@{}}
if not a:\\
\ \ pass
\end{tabular}  &1.82        \\ 
\hline
\begin{tabular}[c]{@{}l@{}}Loop \\Else\end{tabular}               & \begin{tabular}[c]{@{}l@{}}an efficient \\ for loop  imple-\\mentation~\cite{loop_else_performance_so,hettinger2013transforming}\end{tabular} & 
\begin{tabular}[c]{@{}l@{}}
finishedForLoop = True\\
for x in range(2, n):\\
\ \     if n \% x == 0:\\
\ \ \ \         finishedForLoop=False\\
 \ \ \ \        break\\
if finishedForLoop:\\
\ \     pass\\
\end{tabular}             & 
\begin{tabular}[c]{@{}l@{}}
for x in range(2, n):\\
\ \     if n \% x == 0:\\
 \ \ \ \        break\\
else:\\
 \ \     pass\\
\end{tabular}  &
    1.17  \\ 
\hline
\begin{tabular}[c]{@{}l@{}}Assign\\ Multiple\\ Targets\end{tabular}    & \begin{tabular}[c]{@{}l@{}}Speed up and shorten \\your code is to assign \\the variables in your -\\program at the same\\ time~\cite{multi_assign_so,multi_assign_medium,multi_assign_loginradius}\end{tabular} & 
\begin{tabular}[c]{@{}l@{}}
a = 2\\
b = 3\\
c = 5\\
d = 7
\end{tabular}             & 
\begin{tabular}[c]{@{}l@{}}
a, b, c, d \\= 2, 3, 5, 7
\end{tabular}    &1.21 \\ 
\hline
\begin{tabular}[c]{@{}l@{}}Star in \\Func Call\end{tabular}  & \begin{tabular}[c]{@{}l@{}}Unpacking is faster \\than accessing by \\index ~\cite{for_targets_so2,star_so}\end{tabular}  &
\begin{tabular}[c]{@{}l@{}}
s = sum(values[0], \\values[1])
\end{tabular}             & 
\begin{tabular}[c]{@{}l@{}}
s = sum(*values)
\end{tabular}  &1.45         \\ 
\hline
\begin{tabular}[c]{@{}l@{}}For \\Multiple\\ Targets\end{tabular}        & \begin{tabular}[c]{@{}l@{}}Accessing by index\\ slow things down\\ compared to for loop \\item unpacking ~\cite{for_targets_so2,for_targets_so1}\end{tabular} & \begin{tabular}[c]{@{}l@{}}
for item in sales:\\
\ \     a=item[0], item[1], \\ \ \ item[2]

\end{tabular}             & 
\begin{tabular}[c]{@{}l@{}}
for product, price, \\sold\_units in sales:\\
 \ \    a =product, price, \\sold\_units
\end{tabular}   &1.56       \\
\hline

\end{tabular}
\vspace{-2.87em}
\end{table}
\begin{enumerate}[fullwidth,itemindent=0em,leftmargin = 0pt]
    \item[\textbf{RQ1:}]  What are the benefits of pythonic idioms?
	\item[\textbf{RQ2:}] What are the coding practices concerning pythonic idioms and anti-idiom code smells?
	\item[\textbf{RQ3:}]  What are the challenges for writing idiomatic code?
	
\end{enumerate}

\subsection{RQ1: The Benefits of Pythonic Idioms}\label{motiv_2}

Compared with other mainstream programming languages, Python supports more idioms which are highly valued by Python developers~\cite{alexandru2018usage}. 
By referring to several resources (i.e., books, presentations and websites) about pythonic idioms~\cite{alexandru2018usage,reitz2016hitchhiker,knupp2013writing,hettinger2013transforming,slatkin2019effective,merchantepython,StackOverflow,Medium}, we summarize the key benefits of pythonic idioms: conciseness (i.e., fewer lines or fewer tokens) and performance. 
To help readers understand these two benefits, we summarize in Table~\ref{tab:benefits} the code examples of idiomatic code and the corresponding non-idiomatic code (i.e., anti-idiom code smell) for each pythonic idiom excerpted from the reference resources.
We also excerpt relevant performance descriptions.
We omit conciseness related descriptions as code conciseness can be observed intuitively from code examples.

To confirm the performance benefit, we use the \textit{timeit} package~\cite{timeit} to record the execution time of idiomatic and non-idiomatic code snippets shown in Table~\ref{tab:benefits}.
We execute each code snippet three times repeatedly and take the average execution time.
We divide the execution time of non-idiomatic code by that of corresponding idiomatic code which indicates how much speedup idiomatic code has compared with non-idiomatic code.
As shown in the fifth column of Table~\ref{tab:benefits}, idiomatic code has about 1.09\textasciitilde2.07x speedup.
We acknowledge this is only an anecdotal experiment. Many online resources~\cite{list_compr_10X_faster,alexandru2018usage} argue pythonic idioms (e.g., list comprehension) may achieve significant performance advantages over non-idiomatic code, but they generally do not provide specific empirical evidences.
This calls for more systematic investigation of such performance claims which is beyond the scope of this work.

\subsection{RQ2: Python Coding Practices}
\label{sec:challenges}

\subsubsection{Data Preparation}\label{mov_rq1_data}
To understand the coding practices with respect to pythonic idioms and anti-idiom code smells, we crawl the top 10,000 repositories using Python by the number of stars from GitHub. 
7,638 repositories can be successfully parsed using Python 3.
We collect 506,765 Python source files from these repositories.
We then detect the occurrence of idiomatic code and the anti-idiom code that can be refactored with pythonic idioms in these Python source files. 
All nine pythonic idioms can be detected by analyzing abstract syntax trees (ASTs). 
List/set/dict-comprehension and loop-else idioms directly correspond to AST nodes, we can directly detect such idiomatic code instances. 
For star-in-function-call, we extract starred node in function call node. 
For truth-value-test, we extract the test node corresponding to an object. 
For chain-comparison, assign-multiple-targets and for-multiple-targets, we extract operators of the compare node, the value and targets of the assign node, and the target of for node with elements greater than 1, respectively.
For non-idiomatic code, we use our detection rules (see Section~\ref{sec:nonpythonicdetection}) to detect non-idiomatic code instances.
We count the number of repositories, files, methods and statements that contain the instances of idiomatic code and non-idiomatic code. Note that a repository, file or method may contain both pythonic idioms and refactorable non-idiomatic code.

\subsubsection{Result}

\begin{figure}
  \centering

    \includegraphics[width=3.4in]{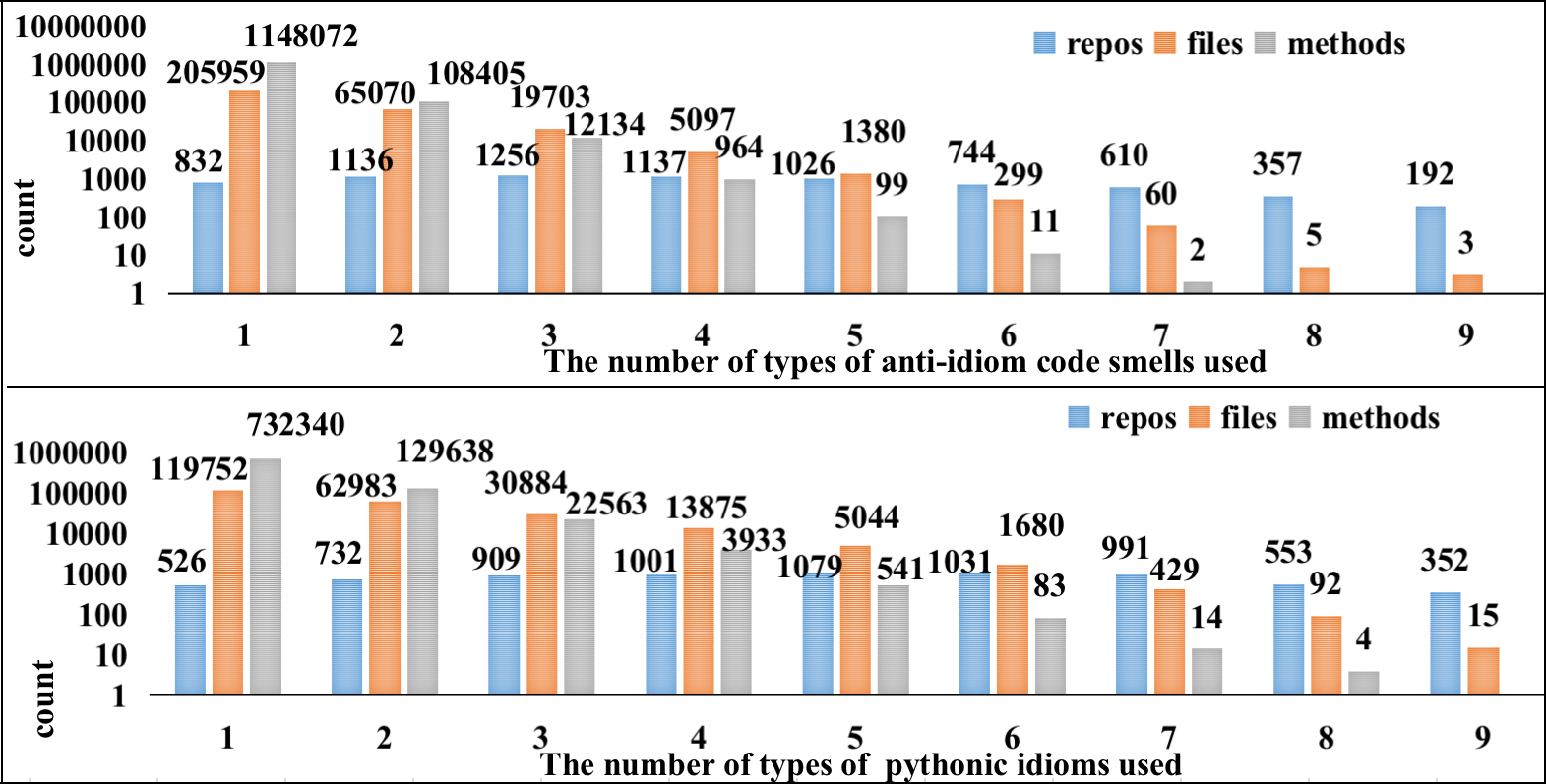}
    \vspace{-0.2cm}
    \caption{The usage of pythonic idioms and anti-idiom code smells in repositories, files and methods}
    \label{fig:nonpythonic_repo_file_met} 
    
\vspace{-0.4cm}
\end{figure}

\begin{table}[tbp]
\scriptsize\caption{Python coding practices with respect to pythonic idioms and anti-idiom code smells}
\vspace{-0.3cm}
  \label{tab:codingpractices}
\centering
\begin{tabular}{|p{0.50in}|p{0.1in} p{0.1in} p{0.19in}|p{0.16in} p{0.15in} p{0.15in}|p{0.18in} p{0.18in} p{0.17in}|} 
\hline
\multirow{2}{*}{Idiom} & \multicolumn{3}{c|}{Repository} & \multicolumn{3}{c|}{File} & \multicolumn{3}{c|}{Method}                                      \\ 
\cline{2-10}
                                & NPy & PyId   & $\bigcap$ 
                                & NPy & PyId   & $\bigcap$ 
                                & NPy & PyId   & $\bigcap$  \\ 
\hline
\begin{tabular}[c]{@{}l@{}}List Compre\end{tabular} & 3814 & 6161  & 3619
&17732&97775&	10040
&24505&	219414&8258	\\ 
\hline
\begin{tabular}[c]{@{}l@{}}Set Compre\end{tabular}      & 700 & 	1048 & 	319	
& 1279 & 	4724 & 151
& 	1512 & 	7151 & 	115  \\ 
\hline
\begin{tabular}[c]{@{}l@{}}Dict Compre\end{tabular}     &  2348	& 3167& 1600& 
7837& 	20871& 	1636& 
10001& 	32941& 1113 \\ 
\hline
\begin{tabular}[c]{@{}l@{}}Chain Compare\end{tabular}    &  4334	 & 2690	 & 2252 & 
26017 & 11066& 	3095
& 39045	& 17241& 2467 \\ 
\hline
\begin{tabular}[c]{@{}l@{}}Truth Val Test\end{tabular}      &  5885	&6930&5728	&
67991&	179184&46303&
135861	&578141&	48550 \\ 
\hline
\begin{tabular}[c]{@{}l@{}}Loop Else\end{tabular} & 806& 	1204& 65& 
1399& 	3602& 133& 
1644& 	4537& 	127  \\ 
\hline
\begin{tabular}[c]{@{}l@{}}Assign Multi\end{tabular}     &7271	&4074&4068&
288578&	28193&	26452&
1173508&	49560&	46075\\ 
\hline
\begin{tabular}[c]{@{}l@{}}Star in Func Call\end{tabular}    & 2336 & 	4185& 1840& 
7671& 	30566& 	1977& 
11325& 	64492& 	2303   \\ 
\hline
\begin{tabular}[c]{@{}l@{}}For Multi\end{tabular}     &  2018 &	5372 &1812	 &
5314 &	57063 &	1953 &
7245 &	104893 &	1923\\
\hline
\end{tabular}
\vspace{-0.4cm}
\end{table}

Figure~\ref{fig:nonpythonic_repo_file_met} shows the numbers of repositories, files and methods using different types of pythonic idioms and anti-idiom code smells. 
We see that although pythonic idioms are well adopted in the Python projects, there are still non-trivial usage of non-idiomatic code that can be refactored with pythonic idioms.
Furthermore, a repository, file or method generally uses multiple types of idiomatic and non-idiomatic code. 
There are non-trivial numbers of repositories and files that use 5 or more types of pythonic idioms or contain 5 or more types of non-idiomatic code that can be refactored with pythonic idioms.


Table~\ref{tab:codingpractices} summarizes the number of repositories, files and methods containing non-idiomatic code smells (the NPy column), pythonic idiom (PyId) and both non-idiomatic and idiomatic code ($\bigcap$) for each pythonic idiom.
We see that many repositories, files or even methods often have a mix of idiomatic and non-idiomatic code to achieve the same functionality.
For example, a large number of repositories and files mix the use of the list-comprehension idiom and non-idiomatic list operation, and 8,258 methods contain both idiomatic list comprehension and non-idiomatic list operation that can be refactored into idiomatic list comprehension.
Among the nine pythonic idioms, set-comprehension and loop-else have relatively low mix usage, while truth-value-test and assign-multiple-targets have high mix usage in methods.

\begin{table}[tbp]
\scriptsize\caption{Statistics of pythonic idioms and anti-idiom code smells at the statement level}
\vspace{-0.3cm}
  \label{tab:codingpractices_code}
\centering
\begin{tabular}{|l|c|c|c|c|} 
\hline
Idiom & Non-idiomatic & Idiomatic   & Sum    &\% \\ 
\hline
List Compre  & 26510  & 349912  & 376422      & 0.070   \\ 
\hline
Set Compre    & 1596  & 9304 & 10900       & 0.146   \\ 
\hline
Dict Compre  & 10695 & 42149 & 52844       & 0.202   \\ 
\hline
Chain Comparison   & 53811 & 26913 & 80724       & 0.667   \\ 
\hline
Truth Value Test   & 197667   & 1023539 & 1221206     & 0.162   \\ 
\hline
Loop Else    & 1644  & 5552  & 7196        & 0.228   \\ 
\hline
Assign Multi Targets  & 1934165 & 70700   & 2004865     & 0.965   \\ 
\hline
Star in Func Call  & 17444    & 90292 & 107736      & 0.162   \\ 
\hline
For Multi Targets & 8490  & 127305 & 135795      & 0.063   \\
\hline
\end{tabular}
\vspace{-0.5cm}
\end{table}

Table~\ref{tab:codingpractices_code} summarizes the occurrence of non-idiomatic code that can be refactored with an idiom (the Non-idiomatic column) and the occurrence of a type of pythonic idiom (the Idiomatic column) at the statement level.
Sum is the sum of the two occurrences and \% is the percentage of the non-idiomatic code out of the Sum.
For the five pythonic idioms (set-comprehension, dict-comprehension, truth-value-test, loop-else and star-in-func-call), the percentages of non-idiomatic code are about 14.6\%-22.9\%.
List-comprehension and for-multi-targets have low non-idiomatic percentages (7.0\% and 6.3\% respectively).
This may be due to the popularity of these two idioms in the Python community. 
Most of online resources we read mention these two idioms.
These two idioms have been used more than 349K and 127K times.
Although the percentage of non-idiomatic code is low, there are still large numbers of non-idiomatic code fragments (26,510 and 8,490) that can be refactored with the list-comprehension idiom and the for-multi-targets idiom respectively.
In contrast, chain-comparison and assign-multi-targets have high non-pythonic percentages (66.7\% and 96.5\% respectively).
Expression comparison and assignment are very basic programming constructs no matter in Python or other programming languages.
However, developers may not realize unique pythonic idioms for comparison and assignment.
For example, developers are surprised when they see Python can make comparison for more than two operands\footnote{\href{https://stackoverflow.com/questions/26502775/simplify-chained-comparison}{https://stackoverflow.com/questions/26502775/simplify-chained-comparison}}.

\subsection{RQ3: Challenges in Writing Idiomatic Code}

We examine Stack Overflow questions to understand the challenges in writing idiomatic code.
We search the questions for each pythonic idiom using the ``python'' tag and the pythonic idiom name.
We examine the returned top 30 questions and summarize the challenges in using pythonic idioms in the discussions.
We summarize three key challenges. 
Table~\ref{tab:motivation_soposts} shows some representative examples.
Many questions have very high view counts which indicate common information needs.
The \#C column lists the challenge index as discussed below. 
One question may involve several challenges.
However, we observe that the three challenges have a progressive relationship.
For example, when developers do not know the meaning of an idiom, they would also further ask how to use the idiom correctly.
Therefore, we list only the most fundamental challenge.

	\textbf{(1) Developers do not know certain pythonic idioms.} 
	For example, for the dict-comprehension idiom (the 4th row in Table~\ref{tab:motivation_soposts}), the developer knows list comprehension but he/she does not know whether he/she can initialize dictionary in a similar way. The question has been viewed about 1,000,000 times. Although it was asked 12 years ago, it was still actively discussed about 1 month ago (as of this paper writing).
	Idioms in Python are more than those in other mainstream languages~\cite{alexandru2018usage}, which brings challenges for developers to learn and write idiomatic Python code. 

	\textbf{(2) Developers know certain pythonic idioms but they do not understand what the idioms can do.} 
	Consider the question for using asterisk operator ``*'' in the function call (the second last row in Table~\ref{tab:motivation_soposts}) which has been viewed about 234,000 times. 
	The developer notices a single asterisk (\textsf{zip(*x)}) can be used before a parameter in function calls, but he/she does not know what this means and what it can be used for. 
Actually, the *x is to unpack x into multiple arguments. Knowing what an idiom can be used for is the pre-requisite for using it in practice.

	\textbf{(3) Developers know what a pythonic idiom can do but they do not know how to use them properly.} 
	The 2nd row of Table~\ref{tab:motivation_soposts} shows such an example.
	The developer wants to refactor a list initialization using list comprehension.
	Unfortunately, he/she does not know whether and how if-else statement can be used in the list-comprehension idiom. 
	The question has been viewed about 165,000 times.
	In fact, list-comprehension has complex syntax and it may nest multiple loops and multiple if statements.
	Developers have to read and understand this complex syntax in order to use the list-comprehension idiom properly. 



\vspace{1mm}
\noindent\fbox{\begin{minipage}{8.4cm} \emph{It has been generally accepted that pythonic idioms result in concise code and improve runtime performance. However, non-idiomatic code that can be refactored with pythonic idioms is widely present in real world projects and is often mix-used with idiomatic code. This could be caused by the fact that developers are often unaware of pythonic idioms or do not know when and how to use pythonic idioms properly. Although there are rich learning materials about pythonic idioms, there are no effective tools to assist developers in writing idiomatic code and to enforce the consistent use of pythonic idioms in practice.} \end{minipage}}\\

\begin{table}[tbp]
\vspace{-0cm}
\scriptsize\caption{Challenges in writing idiomatic Python code}
  \label{tab:motivation_soposts}
  \vspace{-0.4cm}
\centering
\begin{tabular}{|l|c|p{2.4in}|} 
\hline
Idiom          & \#C & Question \\ 
\hline
\begin{tabular}[c]{@{}l@{}}List Com-\\prehension\end{tabular}      & (3)    &\begin{tabular}[c]{@{}l@{}}\href{https://stackoverflow.com/questions/2951701/is-it-possible-to-use-else-in-a-list-comprehension}{\fbox{\textbf{Question: }}}Here is the code I was trying to turn into a list comprehe-\\nsion:...\\Is there a way to add the else statement to this comprehension?\\
            Asked 11 years; Active 2 years; Viewed 165k times\end{tabular}  \\ 
\hline
\begin{tabular}[c]{@{}l@{}}Set Com-\\prehension\end{tabular}       & (1)    & \begin{tabular}[c]{@{}l@{}}\href{https://stackoverflow.com/questions/48421142/fastest-way-to-generate-a-random-like-unique-string-with-random-length-in-python}{\fbox{\textbf{Question: }}}Fastest way to generate a random-like unique string with\\ random length in Python 3\\\textbf{Answer: }...Use a set comprehension to produce a series of keys at a \\time to avoid having to look up and call the set.add() method in a loop...\\
Asked 4 years; Active 6 months ago; Viewed 16k times\end{tabular}            \\ 
\hline
\begin{tabular}[c]{@{}l@{}}Dict Com-\\prehension\end{tabular}      & (1)    & \begin{tabular}[c]{@{}l@{}}\href{https://stackoverflow.com/questions/1747817/create-a-dictionary-with-list-comprehension}{\fbox{\textbf{Question: }}}I like the Python list comprehension syntax. Can it be-\\ used to create dictionaries too? For example, by iterating over pairs of\\ keys and values:\\
Asked 12 years; Active 28 days ago; Viewed 1.0m times\end{tabular}            \\ 
\hline
\begin{tabular}[c]{@{}l@{}}Chain \\Comparison\end{tabular}        & (1)    & \begin{tabular}[c]{@{}l@{}}\href{https://stackoverflow.com/questions/26502775/simplify-chained-comparison}{\fbox{\textbf{Question: }}}I write the following statement: if x \textgreater{}= start and x \textless{}= end:\\ ... the tooltip tells me that I must simplify chained comparison \\What have I missed here?\\\textbf{Comment: }Thanks, I didn't know you could do that in Python.\\ Was really scratching my head on this one.\\Asked 7 years; Active 2 years ago; Viewed 100k times\end{tabular}    \\ 
\hline
\begin{tabular}[c]{@{}l@{}}Truth \\Value\\ Test \end{tabular}       & (1)    & \begin{tabular}[c]{@{}l@{}}\href{https://stackoverflow.com/questions/9573244/how-to-check-if-the-string-is-empty}{\fbox{\textbf{Question: }}}Does Python have something like an empty string \\variable where you can do: if myString == string.empty:
\\Asked 10 years; Active 19 days ago; Viewed 2.5m times\end{tabular}           \\ 
\hline
\begin{tabular}[c]{@{}l@{}}Loop \\Else\end{tabular}               & (2)    & \begin{tabular}[c]{@{}l@{}}\href{https://stackoverflow.com/questions/9979970/why-does-python-use-else-after-for-and-while-loops}{\fbox{\textbf{Question: }}}Why does python use 'else' after for and while loops?\\
Asked 9 years; Active 2 months ago;Viewed 245k times\end{tabular}             \\ 
\hline
\begin{tabular}[c]{@{}l@{}}Assign \\Multiple\\ Targets\end{tabular}    & (1)    & \begin{tabular}[c]{@{}l@{}}
\href{https://stackoverflow.com/questions/16747599/python-assigning-two-variables-on-one-line}{\fbox{\textbf{Question: }}}Python assigning two variables on one line\\
\textbf{Answer: }...use sequence unpacking: self.a, self.b = a, b\\
Asked 8 years; Active 8 years; Viewed 19k times
\end{tabular}     \\ 
\hline
\begin{tabular}[c]{@{}l@{}}Star in \\Func Call\end{tabular} & (2)    & \begin{tabular}[c]{@{}l@{}}\href{https://stackoverflow.com/questions/2921847/what-does-the-star-and-doublestar-operator-mean-in-a-function-call}{\fbox{\textbf{Question: }}}What does the * operator mean in Python, such as in code \\like zip(*x) or f(**k)?\\
Asked 11 years; Active 12 months ago; Viewed 234k times\end{tabular}          \\ 
\hline
\begin{tabular}[c]{@{}l@{}}For \\Multiple\\ Targets\end{tabular}       & (2)    & \begin{tabular}[c]{@{}l@{}}\href{https://stackoverflow.com/questions/10867882/tuple-unpacking-in-for-loops}{\fbox{\textbf{\textbf{Question: }}}}Tuple unpacking in for loops\\
Asked 9 years; Active 1 month ago; Viewed 222k times\end{tabular}            \\
\hline
\end{tabular}
\vspace{-0.4cm}
\end{table}

%% file: method_3_arkiv.tex
\begin{figure}
  \centering
  \includegraphics[width=3.4in]{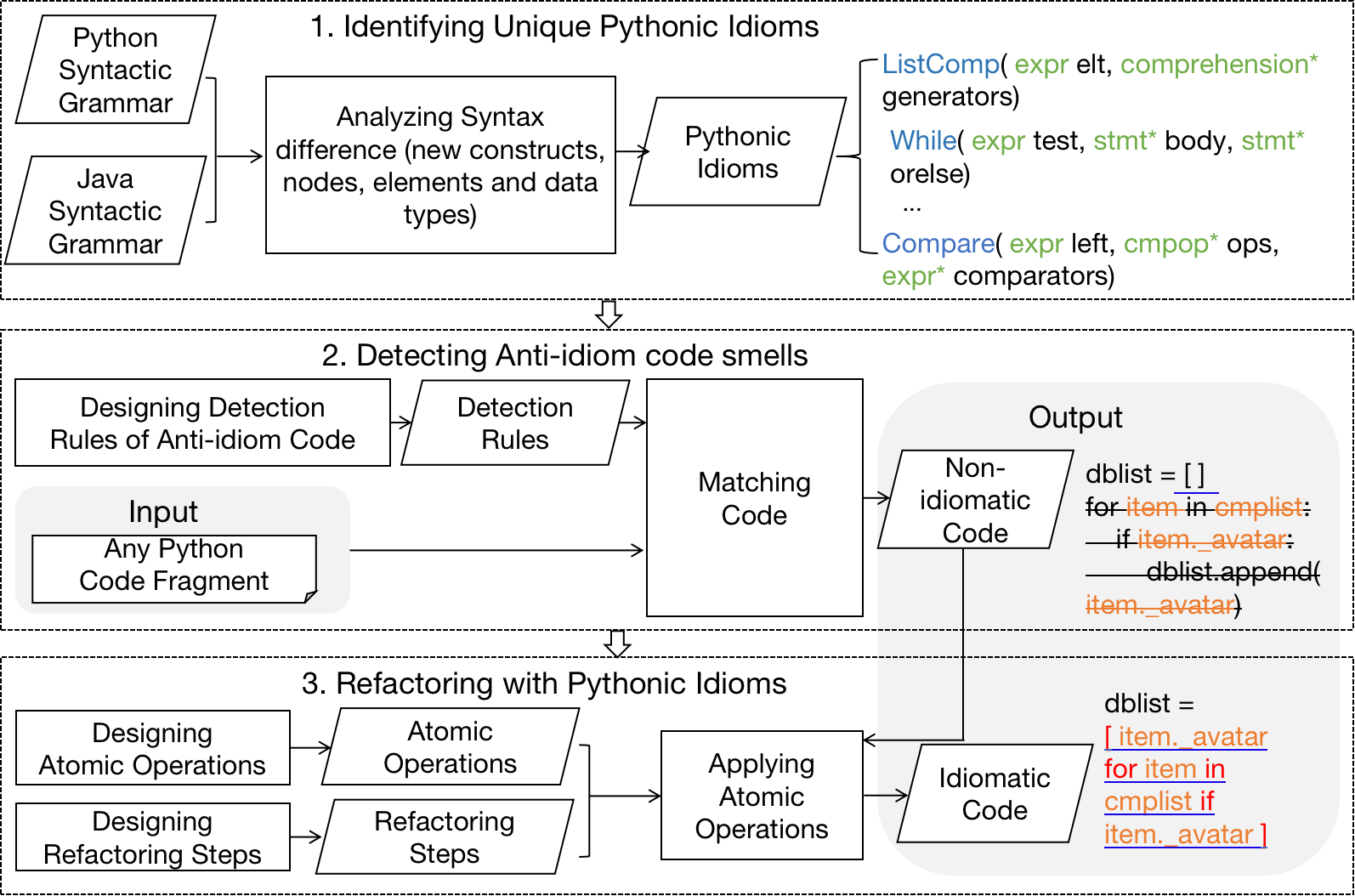}
\setlength{\abovecaptionskip}{-0.2cm}
    \caption{Approach overview}
    
    \label{fig:approach} 
    \vspace{-0.5cm}
\end{figure}

We now present our refactoring tool for improving idiomatic coding practices.
Figure~\ref{fig:approach} shows the three steps for designing and implementing our refactoring tool.
These three steps answer three technical questions respectively: 
1) how to identify programming idioms unique to Python; 
2) how to detect anti-idiom code that can be implemented in pythonic idioms;
3) how to refactor non-idiomatic code with pythonic idioms in a systematic and extensible way.
Rather than relying on mining code patterns or personal programming experience, our solution is built on the effective analysis of Python language syntax and specification.



\subsection{Identifying Unique Pythonic Idioms}\label{idiom}


A programming idiom can be regarded as a micro-level code pattern.
Mining programming idioms in code~\cite{merchantepython,sivaraman2021mining,allamanis2018mining,allamanis2014mining} may identify a wide range of code patterns, including not only pythonic idioms but also many generic code patterns and API usage patterns.
As online materials~\cite{alexandru2018usage,reitz2016hitchhiker,knupp2013writing,hettinger2013transforming,slatkin2019effective,merchantepython} about idiomatic code usually mention only some popular pythonic idioms (e.g., list-comprehension, truth-value-test and enumerate function) repeatedly based on personal programming experience, collecting pythonic idioms from online materials is ad-hoc and may miss important pythonic idioms.

In this work, our goal is to not only identify pythonic idioms but also analyze these idioms to design the methods for detecting and refactoring non-idiomatic code in a systematic manner.
Therefore, we resort to Python language syntax and specification.
We observe that non-idiomatic code share the common or similar syntax as other mainstream programming languages, while pythonic idioms have unique syntax. 
Based on this observation, we identify unique pythonic idioms by contrasting Python syntax and the syntax of the other programming language.
In this work, we choose Java due to its popularity and syntactic similarity to Python.

We examine each syntactic construct of Python in Python language specification~\cite{pythonAPI}, and check whether there is some similar syntactic construct(s) in Java.
For similar syntactic constructs, we further compare and analyze the differences of the syntax grammars between Python and Java. 
We consider the following four cases as candidate pythonic idioms:

\begin{enumerate}[fullwidth,itemindent=0em,label=(1),leftmargin = 0pt]
	\item[(1)] \textbf{Python supports new syntactic constructs which do not exist in Java.} 
	This includes four syntactic constructs: \textsf{ListComp}, \textsf{SetComp}, \textsf{DictComp} and \textsf{Starred} which correspond to four idioms, list comprehension, set comprehension, dictionary comprehension and single asterisk operator. The list/set/dict-comprehension create iterable object with one line of code~\cite{list_compr_pep202,dict_compr_pep274}. 
	The single asterisk operator is usually used to unpack the an iterable into positional arguments inside a function call (star-in-func-call)~\cite{star_pep_448}. 
	We also identify several other new constructs, such as \textsf{Yield}, \textsf{With} and \textsf{GeneratorExp}.
	However, it is very inconvenient to implement the same functionality as these new constructs in a non-idiomatic way.
	As such, the \textsf{Yield}-, \textsf{With}- or \textsf{GeneratorExp}-equivalent non-idiomatic code is too complex to safely refactor.
	Therefore, we do not consider \textsf{Yield}, \textsf{With} and \textsf{GeneratorExp} in this work.

	\item[(2)]\textbf{Python syntax adds new nodes.} 
	Python and Java have the same syntactic construct, but Python syntax adds new nodes to extend its functionality. 
	This includes the \textsf{Loop} construct which consists of the \textsf{For} and \textsf{While} statement. 
	For example, the \textsf{For} statement of Python adds \textsf{orelse} node (i.e., the loop-else idiom). 
	The \textsf{orelse} node is executed after the loop iterator is exhausted, unless the loop ends prematurely due to a break statement~\cite{knupp2013writing}.

	\item[(3)] \textbf{Python syntax adds new elements.} 
	Python and Java have the same syntactic construct, but Python syntax supports more elements~\cite{pythonAPI}. 
	This includes the \textsf{Assign} statement with multiple targets (the assign-multi-targets idiom), the \textsf{ChainComp} with multiple operators (the chain-comparison idiom), and the \textsf{For} statement with multiple targets (the for-multi-targets idiom). 
	For example, the assignment statement of Python allows multiple variables to be assigned simultaneously. 
	A useful scenario for assign-multi-targets is to swap variables without creating temporary variables. 
	
	\item[(4)] \textbf{Python syntax supports more comprehensive data type.} 
	Python and Java have the same syntactic construct, but Python supports objects of more data types. 
	This includes the truth-value-testing idiom~\cite{truth_value_test_python_doc,knupp2013writing}. 
	In Python, any object (e.g., string, numeric type and sequences) can be directly tested for truth value. 
	For example, we can directly check if a variable \textsf{``a''} of list data type is empty with \textsf{``if not a''} instead of \textsf{``if a == []''}.
\end{enumerate}

Table~\ref{tab:benefits} lists the nine pythonic idioms we identify and the code examples of idiomatic code and corresponding non-idiomatic code.
For each identified pythonic idiom, we read Python language specification to confirm its validity.
We also search online materials with the idiom names to support our analysis.
Searching with specific idiom names can find relevant supporting documentation.
However, searching ``python idiom'' generally return many materials which do not cover all nine types of pythonic idioms.

\begin{table*}[]
 \footnotesize
  \caption{Examples of detection and refactoring of anti-idiom code smells}
  \vspace{-0.4cm}
  \label{refactor_rule}
\centering
\begin{tabular}{|l|@{}c@{}|l|}
\hline
Idiom   & \multicolumn{1}{c|}{Detection Rules and Examples of Code Pairs} & \multicolumn{1}{c|}{Refactoring Steps} \\ \hline

\begin{tabular}[c]{@{}l@{}}
\includegraphics[width=0.37in]{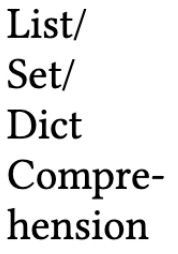}\end{tabular} &
\begin{tabular}[c]{@{}l@{}}
\includegraphics[width=4.0in]{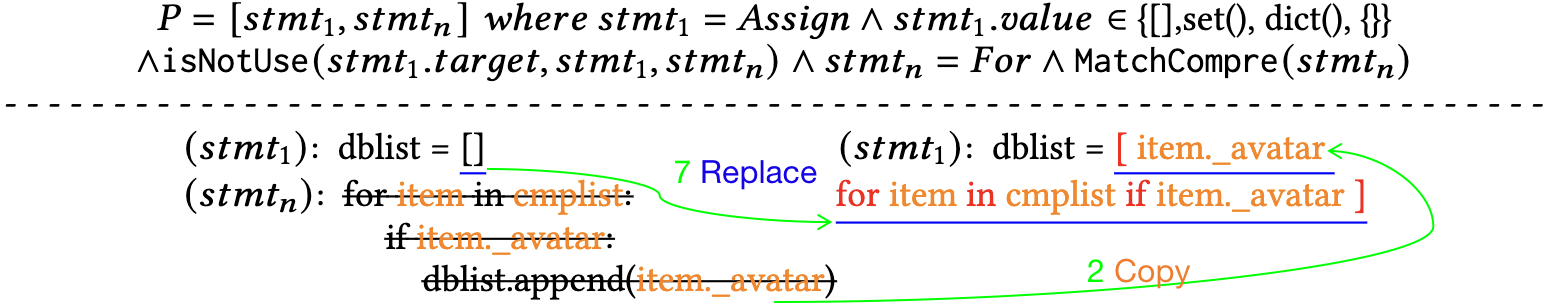}
 \end{tabular}
&
\begin{tabular}[l]{@{}l@{}}
\includegraphics[width=2.2in]{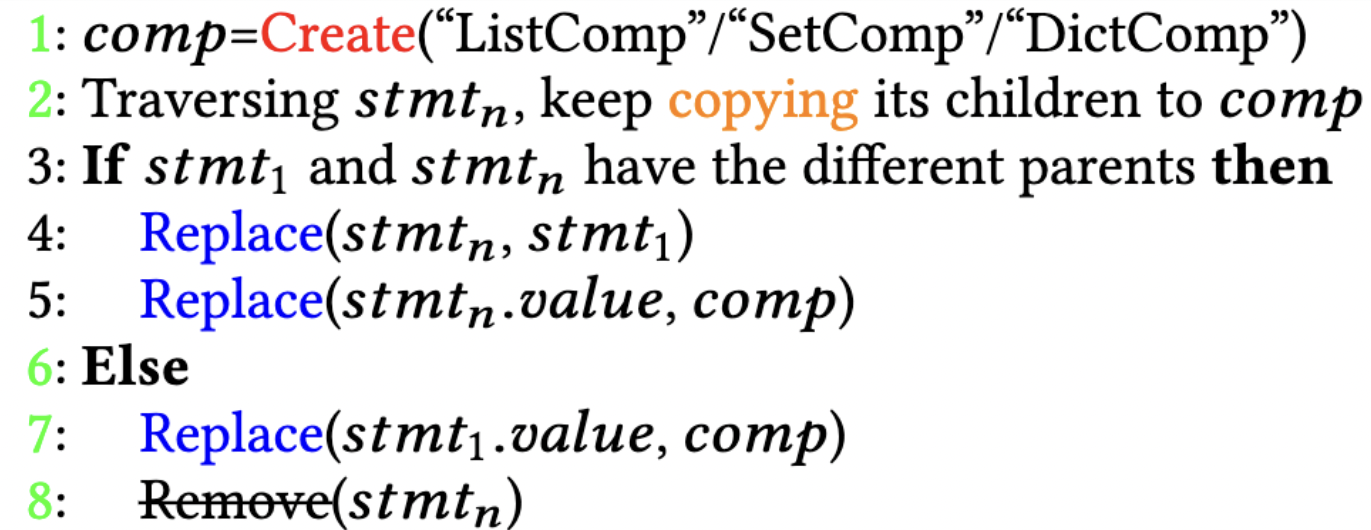}
\end{tabular}         \\ 
\hline
\begin{tabular}[c]{@{}l@{}}\includegraphics[width=0.5in]{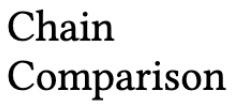}
\end{tabular}        &\begin{tabular}[c]{@{}l@{}}
\includegraphics[width=4.0in]{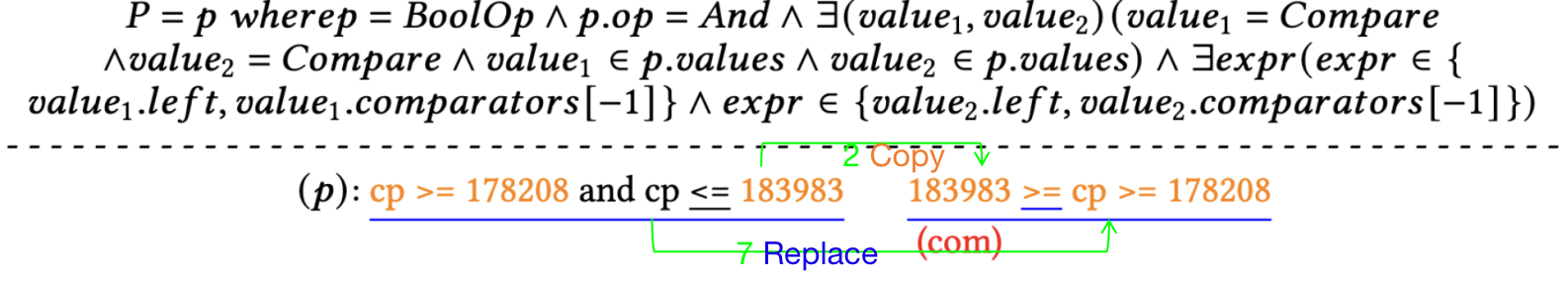}
        \end{tabular}
        &
        \begin{tabular}[c]{@{}l@{}}
\includegraphics[width=2.2in]{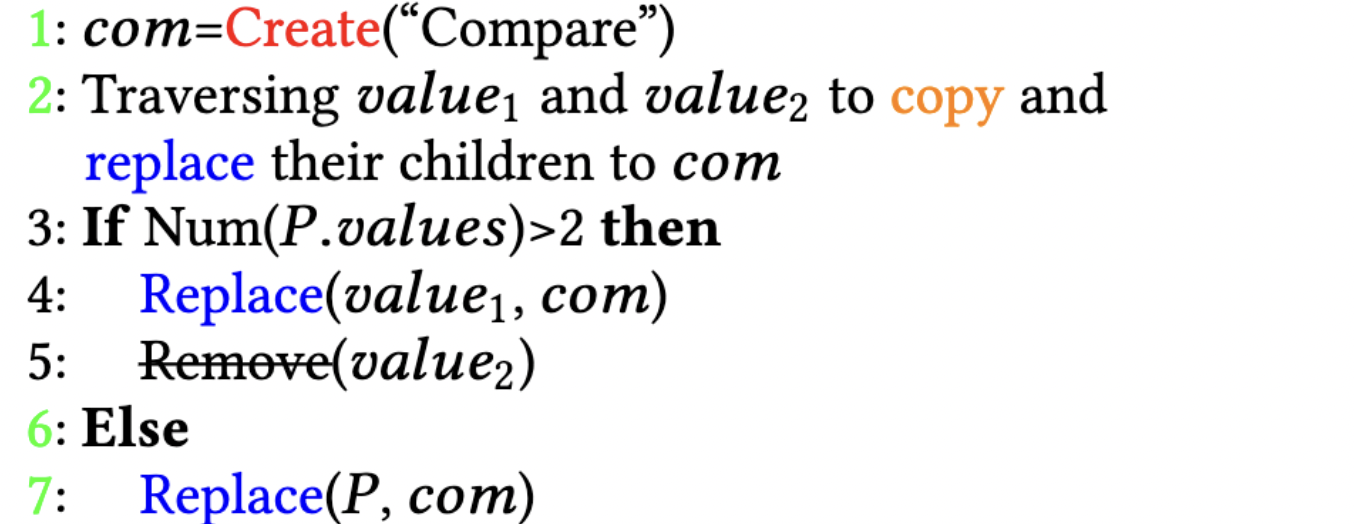}
\end{tabular}\\ 
     \hline
\begin{tabular}[c]{@{}l@{}}
\includegraphics[width=0.28in]{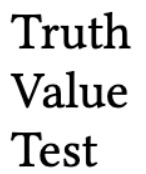}\end{tabular}        &
\begin{tabular}[c]{@{}l@{}}
\includegraphics[width=4.0in]{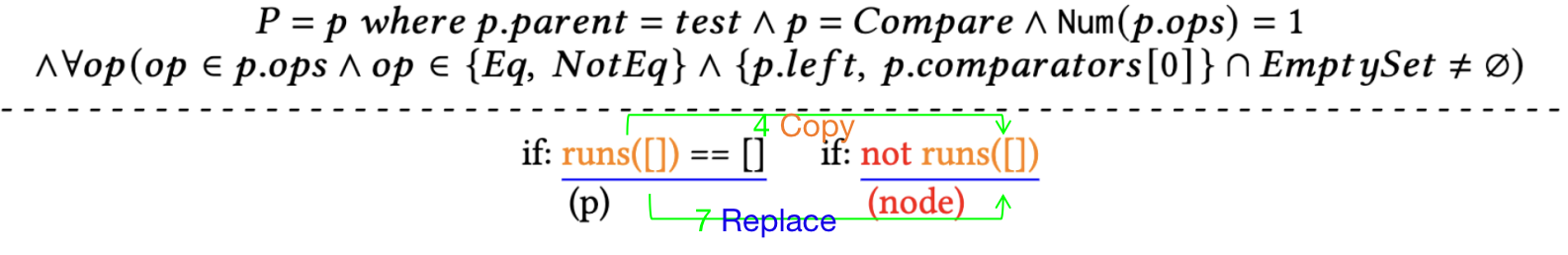}
\end{tabular}&
\begin{tabular}[c]{@{}l@{}}
\includegraphics[width=2.2in]{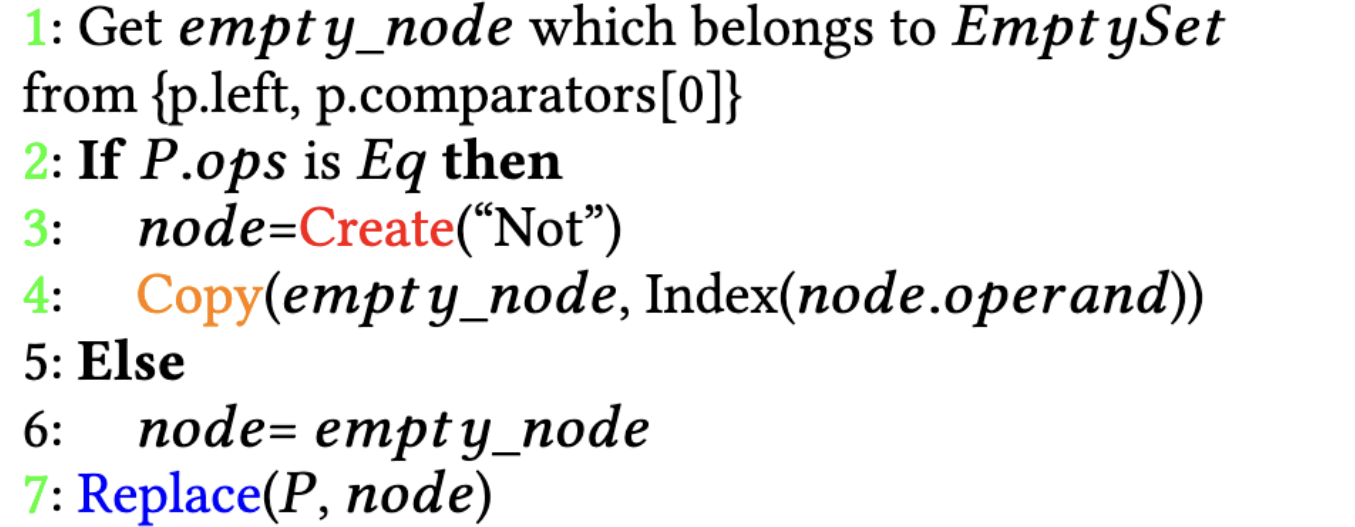}
\end{tabular}         \\ 
\hline

\begin{tabular}[c]{@{}l@{}}\includegraphics[width=0.23in]{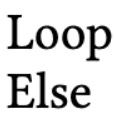}\end{tabular}        &
\begin{tabular}[c]{@{}l@{}}
\includegraphics[width=4.1in]{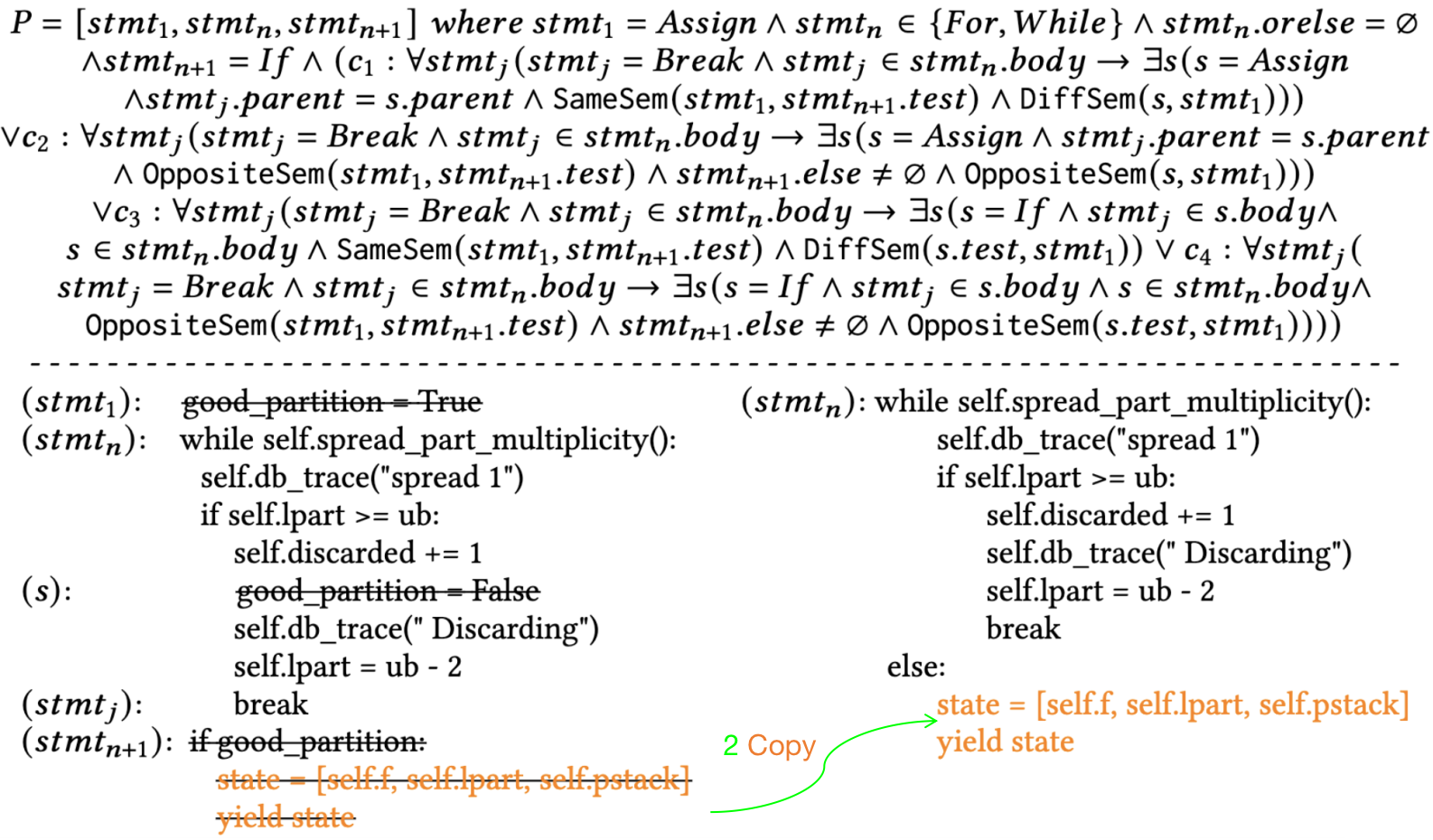}

\end{tabular}&
\begin{tabular}[c]{@{}l@{}}
\includegraphics[width=2.1in]{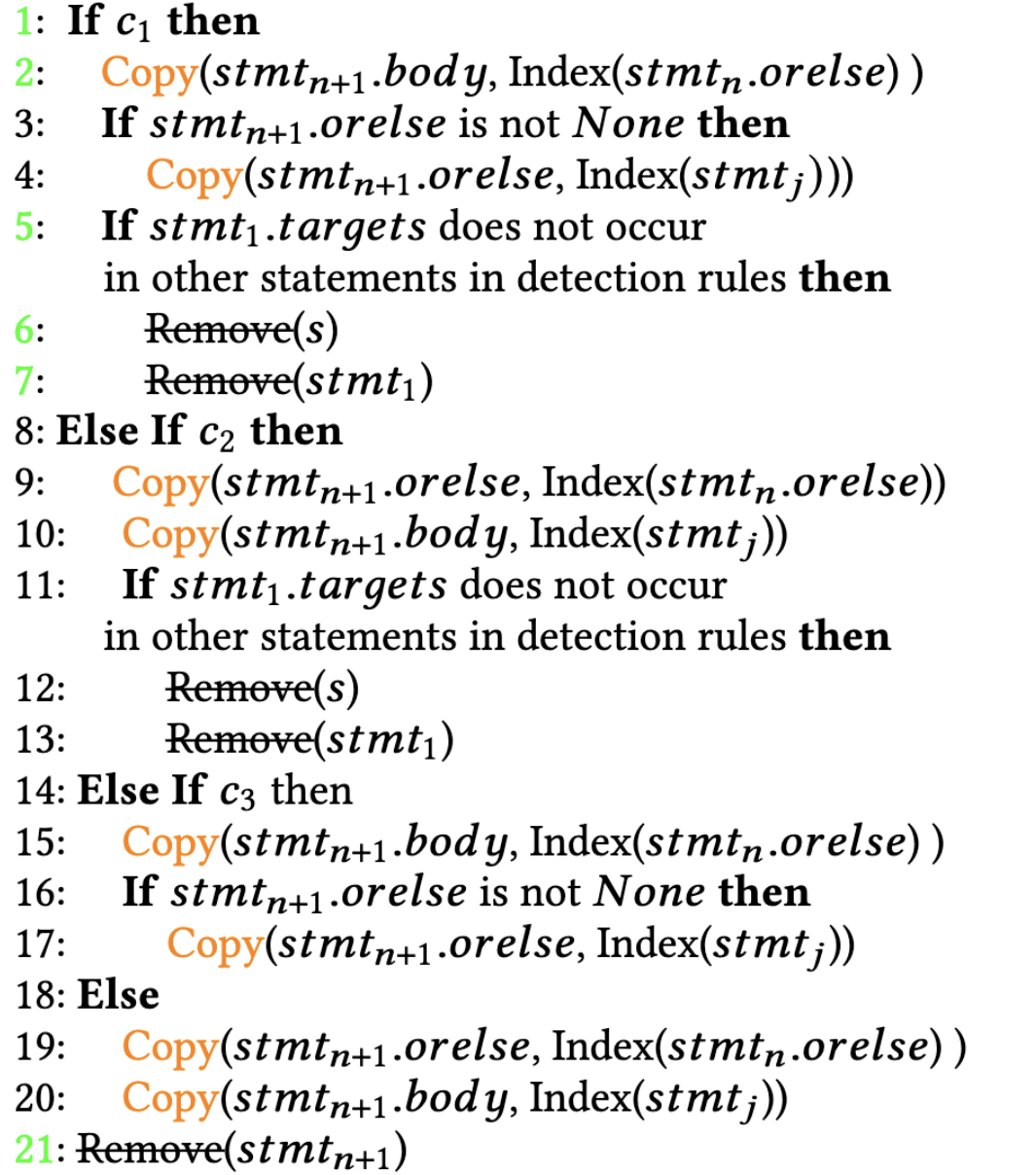}
    \end{tabular}      \\ 
\hline

\begin{tabular}[c]{@{}l@{}}\includegraphics[width=0.34in]{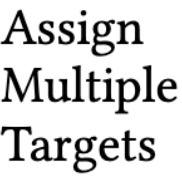}
\end{tabular}        &
\begin{tabular}[c]{@{}l@{}}
\includegraphics[width=3.95in]{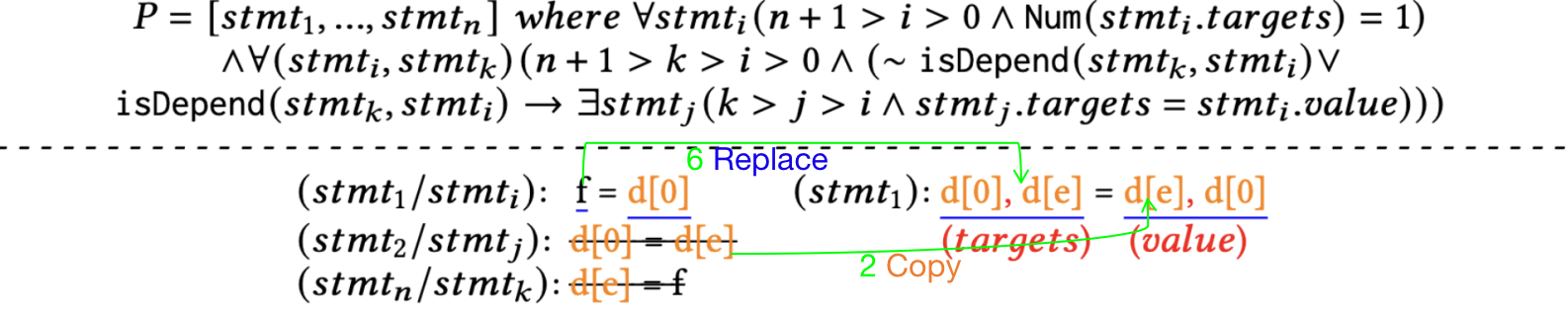}
\end{tabular}  &
\begin{tabular}[c]{@{}l@{}}
\includegraphics[width=2.0in]{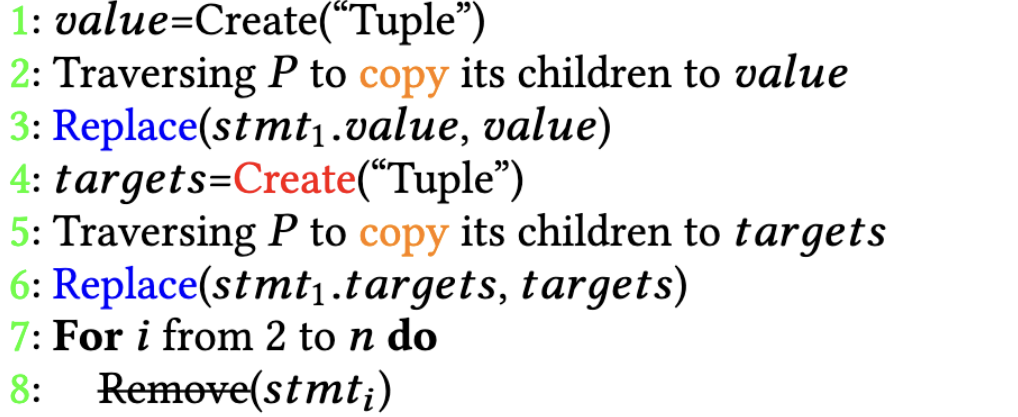}
\end{tabular}  
     \\
     
\hline
\begin{tabular}[c]{@{}l@{}}\includegraphics[width=0.23in]{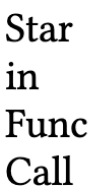}
\end{tabular}        &\begin{tabular}[c]{@{}l@{}}
\includegraphics[width=3.8in]{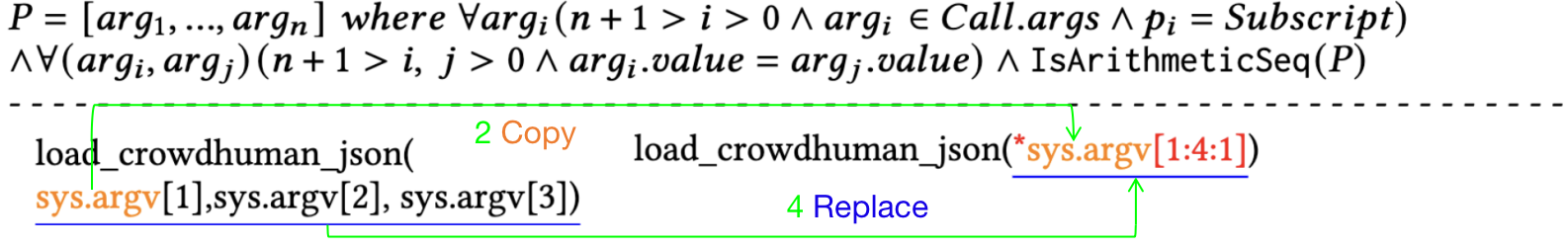}
        \end{tabular}
        &
        \begin{tabular}[c]{@{}l@{}}
\includegraphics[width=2.2in]{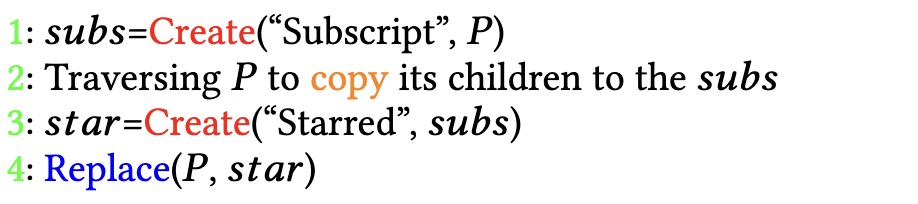}
\end{tabular}\\ 
     \hline
\begin{tabular}[c]{@{}l@{}}\includegraphics[width=0.37in]{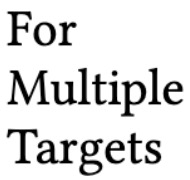}
\end{tabular}        &\begin{tabular}[c]{@{}l@{}}
\includegraphics[width=3.8in]{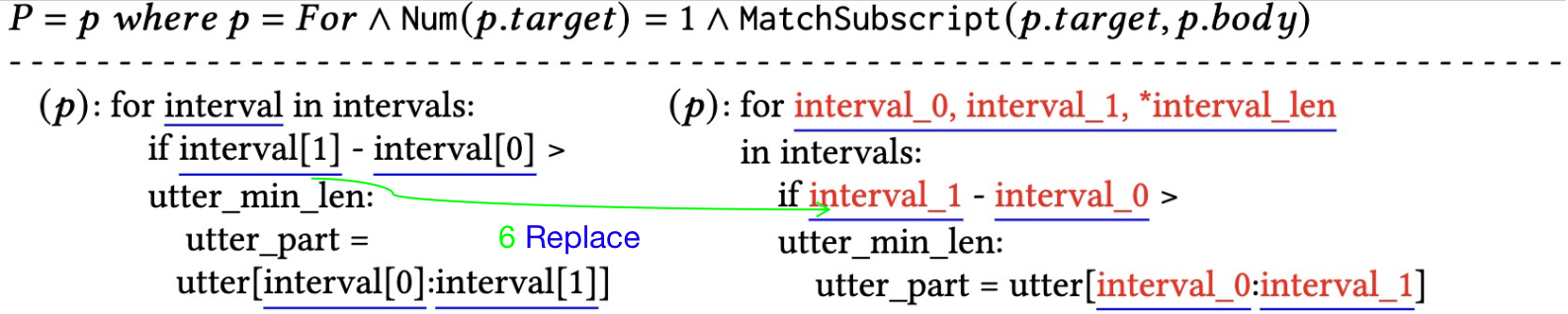}
        \end{tabular}
        &
        \begin{tabular}[c]{@{}l@{}}
\includegraphics[width=2.2in]{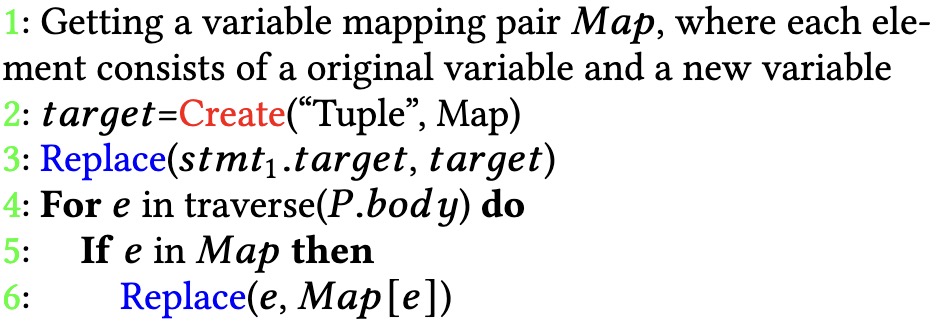}
\end{tabular}\\ 
     \hline
\end{tabular}
 {\raggedright 
\includegraphics[width=7.0in]{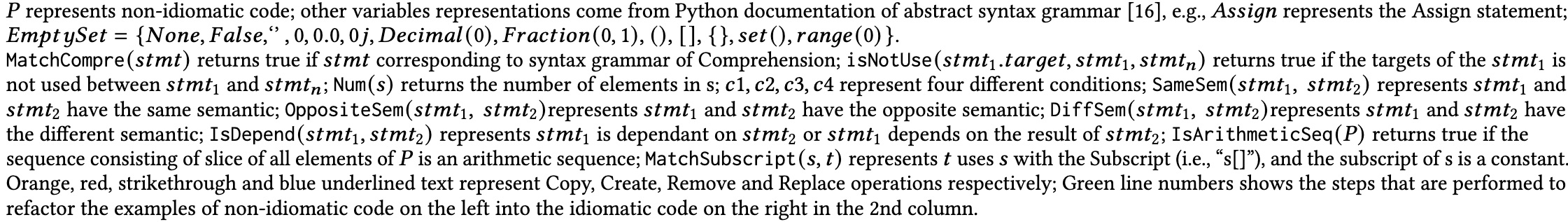} 

 \par}
\end{table*}
\subsection{Detecting Anti-idiom Code Smells}\label{sec:nonpythonicdetection}

For each pythonic idiom, we define syntactic patterns for implementing the pythonic idioms in a non-idiomatic way.
Contrasting Python and Java syntax provides the basis for defining such patterns.
In a sense, we try to implement a pythonic idiom in a Java-style Python code pattern.
The defined syntactic patterns can detect anti-idiom code smells that can be refactored with pythonic idioms.
Table~\ref{refactor_rule} lists our detection rules and illustrative examples.


\subsubsection{List/Set/Dict Comprehension}

The list-comprehension idiom is used for the list initialization (2nd row in Table~\ref{refactor_rule}). 
The rule first finds an empty assignment statement $stmt_1$ (e.g., \textsf{``dblist = []''}). 
Then, it finds a \textsf{for} statement $stmt_n$ which iteratively adds elements to the target (\textsf{ ``dblist''}) of $stmt_1$. There cannot be other statements using the target \textsf{``dblist''} of $stmt_1$ between $stmt_1$ and $stmt_n$ to lest the \textsf{``dblist''} is modified (i.e., $isNotUse(stmt_1.target, stmt_1, stmt_n)$). 
Since the $stmt_n$ corresponds to the \textsf{comp} node of the \textsf{ListComp} construct which only supports \textsf{for} clause and \textsf{if} clause, the rule checks whether $stmt_n$ satisfies the \textit{MatchCompre} condition, i.e., if the $stmt_n$ corresponds to the syntax grammar of Comprehension. 
The detection rule for the non-idiomatic code of the set-comprehension and the dict-comprehension idiom are the same.

\subsubsection{Chain Comparison}
The chain-comparison ``a $op_1$ b $op_2$ c ... y $op_n$ z'' is equivalent to ``a $op_1$ b and b $op_2$ c and ... y $op_n$ z''~\cite{chainCompare}. 
The non-idiomatic code of the chain comparison must be a $BoolOp$-\textsf{and} expression which contains at least two compare nodes. 
Moreover, the two compare nodes have the same operands. 
For example, for the expression ``\textsf{cp >= 178208 and cp <= 183983}'' (3rd row in Table~\ref{refactor_rule}), the \textsf{cp} is the common operand of the two compare nodes, and the expression can be refactored as \textsf{``183983 >= cp >= 178208''}.

\subsubsection{Truth Value Test}
The truth-value-test idiom is used for checking the ``truthiness'' of an object. 
Generally, when developers want to compare whether an object is equal or is not equal to a value, many programming languages use ``=='' or ``!='' operator to achieve the functionality. 
In Python, any object can be directly tested for truth value, so developers do not need to use ``=='' or ``!='' operator to test truth value. 
Python documentation specify the built-in objects in $EmptySet$ (e.g., \textsf{[]} and \textsf{set()}) are considered as false value.
Therefore, if a statement directly compares an object to the element of $EmptySet$, it will be regarded as a non-idiomatic code of the truth value test. 
However, not all compare nodes are refactorable with truth value test. 
For example, \textsf{``a!=[]''} in \textsf{``return a!=[]''} cannot be refactored because ``\textsf{return a}'' changes the code semantic.
According to Python syntax, the non-idiomatic code of truth value test corresponds to a \textsf{test}-type node. 
Therefore, our rule checks whether a compare node is the child of a \textsf{test}-type node, for example, the \textsf{``runs([]) == []''} is the child of an \textsf{if}-node \textsf{``if runs([]) == []''} (4th row in Table~\ref{refactor_rule}). 
Since the \textsf{if}-node is a \textsf{test}-type node, the compare node \textsf{``runs([]) == []''} is refactorable to a truth-value-test.

\subsubsection{Loop Else}
The else clause of the loop statement is executed after the iterator is exhausted, unless the loop was ended prematurely due to a break statement. The non-idiomatic way of implementing a loop-else generally has an assignment statement $stmt_1$ to flag current state, a \textsf{for} statement $stmt_n$ which contains a statement $s$ to change the current state and a \textsf{break} statement $stmt_j$ to end the loop,  and an \textsf{if} statement $stmt_{n+1}$ after the \textsf{for} statement $stmt_n$ to check the current state to execute different operations. There are four circumstances: $c_1$ and $c_2$ complement each other, and $c_3$ and $c_4$ complement each other. 

The \textbf{$c_1$} satisfies the following semantic conditions: the semantic of the assignment statement $stmt_1$ is the same as the semantic of the test node of if statement $stmt_{n+1}.test$, and the semantic of assignment statement $s$ is different from the semantic of $stmt_1$ where $s$ and the break statement $stmt_j$ are at the same scope.
These semantic conditions are designed because the non-idiomatic code of loop-else implies two execution paths (5th row in Table~\ref{refactor_rule}): $stmt_1$ $\nrightarrow$ $s$ and $stmt_1 \rightarrow stmt_{n+1}$ or $stmt_1 \rightarrow s$ and $stmt_j \nrightarrow stmt_{n+1}$.

The $c_2$ satisfies the following semantic conditions: the semantic of the assignment statement $stmt_1$ is the opposite of the semantic of the test node of the if-statement $stmt_{n+1}.test$, the if-statement $stmt_{n+1}$ has an else clause, and the semantic of the assignment statement $s$ is the opposite of the semantic of $stmt_1$ where $s$ and the break statement $stmt_j$ are at the same scope. 
The $c_2$ condition is a complement to the $c_1$ condition. 
If $stmt_{n+1}$ has an else clause and $stmt_{n+1}.test$ has the opposite semantic with $stmt_1$, it indicates that the else clause has the same semantic as $stmt_1$. 
Therefore, the code satisfying the $c_2$ condition is also refactorable to a loop-else. 
For example (5th row in Table 5), if we change $stmt_{n+1}.test$ \textsf{``good\_partition''} into \textsf{``not good\_partition''} and add an else clause to the if statement, the code satisfies the $c_2$ condition.  

The $c_3$ satisfies the following semantic conditions: 
the semantic of the assignment statement $stmt_1$ is the same as the semantic of test node of the if-statement $stmt_{n+1}.test$, and the semantic of the if-statement $s$ in the body of the loop statement $stmt_{n}$ is different from the semantic of $stmt_1$, and the body of the if-statement $s$ contains the break statement $stmt_j$. 
The $c_3$ is a variant of $c_1$ and $c_2$. 
The $c_1$ and $c_2$ requires an assignment $s$ to change the current state, but $c_3$ uses an if statement $s$ to detect the change of the current state and break the loop, such as ``\textsf{if not good\_partition: break}''. 

The $c_4$ satisfies the following semantic conditions: the semantic of the the assignment $stmt_1$ is the opposite of the semantic of test node of the if-statement $stmt_{n+1}.test$, the if-statement $stmt_{n+1}$ has an else clause, and the semantic of the test node of if statement $s.test$ in the body of the loop-statement $stmt_{n}$ is the opposite of the semantic of $stmt_1$ and the body of the if-statement $s$ contains the break statement $stmt_j$. 
The $c_4$ complements $c_3$, in the same vein as $c_2$ complements $c_1$.

\subsubsection{Assign Multiple Targets}
The assign-multiple-targets idiom is to assign multiple values at the same time in one assignment statement. 
For several consecutive assignment statements, if an assignment statement $stmt_k$ does not use the result of an assignment statement $stmt_i$ before it, these assignment statements are refactorable to assign-multi-targets. 
When an assignment statement $stmt_k$ uses the result of the an assignment statement $stmt_i$ before $stmt_k$, the code usually is to swap variables by creating temporary variables. 
For such non-idiomatic code, it requires that the target of a statement $stmt_j$ between the $stmt_i$ and the $stmt_k$ is the same as the value of $stmt_i$. 
For example (6th row in Table~\ref{refactor_rule}), $stmt_k$ \textsf{``d[e] = f''} uses the target \textsf{``f''} of $stmt_i$ \textsf{``f = d[0]''}, and the target \textsf{``d[0]''} of the $stmt_j$ \textsf{``d[0] = d[e]''} is the same as the value \textsf{``d[0]''} of the $stmt_i$ \textsf{``f = d[0]''}.
This sequence of assignments via a temporary variable can also be refactored with the assign-multiple-targets idiom.

\subsubsection{Star in Function Calls}
The star-in-function-call idiom is usually used to unpack an iterable to the positional arguments in a function call~\cite{star_pep_448}. 
The non-idiomatic way of passing a sequence of arguments is that the subscript sequence of multiple consecutive parameters of a function call is an arithmetic sequence of the same variable. 
For example, ``1, 2, 3'' is an arithmetic sequence where the common difference is 1 for accessing the first, second and third element of \textsf{``sys.argv''} (second-to-last row in Table~\ref{refactor_rule}). It can be refactored into “*sys.argv[1:4:1]”.

\subsubsection{For Multiple Targets}
The non-idiomatic code of the for-multiple-targets idiom only contains one variable as the target of for statement $p$. 
The body of $p$ uses the subscript expression to get elements of the variable. 
For example (the last row of Table~\ref{refactor_rule}), the code uses \textsf{``interval[0]''} and \textsf{``interval[1]''} to get the elements of the variable \textsf{``interval''} inside the body of for loop.
Instead, the elements of \textsf{``interval''} can be accessed using a for-multiple-targets idiom.

\subsection{Refactoring with Pythonic Idioms}\label{refactor}

According to~\cite{fowler2018refactoring}, a refactoring is a series of small behavior preserving transformations.
Based on this principle, we analyze the AST transformations required to transform a piece of anti-idiom code into an idiomatic code.
We identify four atomic AST-rewriting operations across all idioms, and then compose these atomic operations into the refactoring steps for each pythonic idiom. 
The four atomic operations are as follows:

\begin{enumerate}[fullwidth,itemindent=0em,label=(1),leftmargin = 0pt]
 	\item[(1)] \textbf{Copy(s, i)} copies the node $s$ of non-idiomatic code to the position $i$ of a node of idiomatic code. If the node at the position $i$ is empty, we copy $s$ into the position $i$. Otherwise, we insert $s$ into the position $i$. Since a refactoring does not change the code semantics, many parts of non-idiomatic code can be copied to the resulting idiomatic code. 
 	For example, for the list-comprehension idiom (2nd row in Table~\ref{refactor_rule}), both the target node \textsf{item} and the iter node \textsf{cmplist} of non-idiomatic code are copied to the corresponding target and iter position of the comprehension node respectively. For another example, for the chain-comparison idiom (3rd row in Table~\ref{refactor_rule}) we copy operands of \textsf{compare} node of non-idiomatic code into the position of operands of a new \textsf{compare} node.

 	\item[(2)]\textbf{Create(s, *info)} builds the node of type $s$ with information $*info$ where * represents any amount of information. To refactor non-idiomatic code into pythonic idioms, it is sometimes necessary to create some new AST nodes or elements which do not have the corresponding parts in the non-idiomatic code. For example, for the truth-value-test idiom (4th row in Table~\ref{refactor_rule}), we need to create a \textsf{``Not''} node. For another example, for the star-in-function-call idiom (second-to-last row in Table~\ref{refactor_rule}), we need to create a \textsf{Starred} node with subscript information from the non-idiomatic code.

	\item[(3)] \textbf{Remove(s)} removes the node $s$ from the AST of non-idiomatic code which is no longer needed in idiomatic code. Generally, refactoring non-idiomatic code into idiomatic code will reduce the lines or tokens of code. Therefore, it is natural to remove those no-longer-used nodes. For example, for the loop-else idiom (5th row in Table~\ref{refactor_rule}), we need to remove the initial flag assignment \textsf{``good\_partition = True''} and the flag-update statement \textsf{``good\_partition = False''} which are no longer needed when the loop-else idiom is used. For another example, for the assign-multi-targets idiom (6th row in Table~\ref{refactor_rule}), we remove assign statements from $stmt_2$ to $stmt_n$.
	
	\item[(4)] \textbf{Replace(s, t)} replaces the node $s$ of non-idiomatic with the node $t$ obtained through code transformation.
	For example, for the chain-comparison idiom (3rd row in Table~\ref{refactor_rule}), we replace the original expression \textsf{``cp >= 178208 and cp <= 183983''} with the resulting chain-comparison \textsf{``183983 >= cp >= 178208''}. For another example, for the for-multi-targets idiom (the last row in Table~\ref{refactor_rule}), we replace \textsf{``interval[0], interval[1]''} with \textsf{``interval\_0, interval\_1''} respectively.
	
\end{enumerate}

The 3rd column of Table~\ref{refactor_rule} shows the refactoring steps to complete each pythonic idiom refactoring. 
The green line numbers shows the steps that are performed to refactor the examples of non-idiomatic code on the left into the idiomatic code on the right in the 2nd column of Table~\ref{refactor_rule}. 
For example, to refactor the non-idiomatic code example into a list comprehension code (2nd row in Table~\ref{refactor_rule}), we first create a ListComp node $comp$ and then traverse the \textsf{for} statement $stmt_n$ to copy its children to the $comp$ node (line 1-2), e.g., copy \textsf{item.\_avatar} to the position of $stmt_n.elt$ (i.e., elements to add to the list). Since $stmt_n$ and $stmt_1$ are at the same scope (line 6), we directly replace $stmt_1.value$ with $comp$ in and then remove $stmt_n$ (line 7-8). 
Finally, the new $stmt_1$ is the idiomatic code obtained through the refactoring. When $stmt_1$ and $stmt_n$ are at different scope (line 3), we do not perform the Remove operation for the $stmt_1$ because $stmt_n$ may not be executed after executing $stmt_1$, so we only replace $stmt_n$ with $stmt_1$ and then update the value of $stmt_n$ (line 4-5).

%% file: result_3_arkiv.tex
This section reports the evaluation of our approach.
We focus on two aspects: the correctness and usefulness of refactoring anti-idiom code smells with pythonic idioms: 

\begin{enumerate}[fullwidth,itemindent=0em,leftmargin = 0pt]
	\item[\textbf{RQ1:}] How accurate is our approach when refactoring real-world anti-idiom Python code with pythonic idioms? 
	\item[\textbf{RQ2:}] Do code refactorings performed by our approach have practical value for real-world projects? 
\end{enumerate}
\vspace{-0.1cm}	
\subsection{RQ1: Correctness of Refactorings}\label{rq1}

\subsubsection{Motivation}
Refactorings involve complex logic for detecting anti-idiom code smells and applying code transformation.
We would like to confirm the design and implementation of our approach is robust and correct on real-world Python code.

\subsubsection{Method}
As described in Section~\ref{mov_rq1_data}, we collect 7,638 repositories from GitHub which can be successfully parsed using Python 3, and collect 506,765 Python source files from these repositories.
We apply our refactoring tool to these Python source files to detect nine types of anti-idiom code statements and refactor these statements with pythonic idioms.
We use both testing and code review to verify the correctness of refactorings.

\textbf{Testing based verification.}
To determine the test cases that cover the detected  non-idiomatic code fragments, we first collect the fully qualified names of all methods called by a test method using the DLocator tool~\cite{wang2020exploring}.
If the method that contains a  non-idiomatic code fragment belongs to the list of the methods called by the test method, we consider this test method as a test case for the  non-idiomatic code fragment. Note that one test case may test one or more methods, and one method may undergo one or more different types of refactorings. Then, to execute the test cases successfully, we install the packages that the project depends on by reading its requirements.txt. 
Note that not all test cases can be executed successfully because of several problems, such as requiring other non-python packages or to manually configure some parameters. 
We filter out such fail-to-execute test cases.

In the work, we use Pytest~\cite{hunt2019pytest}, a popular Python unit testing frame which also supports the Python's default unittest tool~\cite{unittest}.
We run the test cases on the original methods with  non-idiomatic code fragments to ensure they pass successfully.
Then we run the test cases again on the refactored methods.
If the refactored methods pass the test cases, we consider the detection of anti-idiom,  non-idiomatic code fragments and the corresponding code refactorings are correct.
For the refactorings that fail the test cases, the two authors independently analyze the failure causes.
Two authors have more than three years of Python development experience.
They examine the detected  non-idiomatic code fragments and the idiomatic code obtained by the refactorings, and determine if the failures are caused by  non-idiomatic code smell detection or pythonic idiom transformation. 
A detection failure means a detected  non-idiomatic code fragment is not refactorable, e.g, it cannot be safely refactored into semantic-equivalent idiomatic code. 
If the failure is caused by  non-idiomatic code detection, we do not double count it as the failure of pythonic idiom transformation. 
The two authors discuss to resolve their disagreement and reach the consensus.
Finally, we compute the accuracy of anti-idiom code smell detection and idiomatic code refactoring for each pythonic idiom.

\textbf{Code review based verification}.
We randomly sample 100 pairs of  non-idiomatic code fragments and the corresponding idiomatic code fragments for each pythonic idiom. 
Then the two authors independently review these code pairs, and determine if the  non-idiomatic code fragments are detected correctly and if the idiomatic code fragments are refactored correctly. 
They discuss to resolve their disagreement and reach the consensus.
Based on their code review results, we compute the accuracy of anti-idiom code smell detection and idiomatic code refactoring for each pythonic idiom.
\begin{table}[htbp]
\footnotesize
\caption{Accuracy of anti-idiom code smell detection (d-acc) and idiomatic code transformation (r-acc)}
  \label{tab:correctness}
\vspace{-0.4cm}
\centering
\begin{tabular}{|l|c|c|c|c|c|c|c|} 
\hline
\multirow{2}{*}{Idiom}                                    & \multicolumn{4}{c|}{Testing}                & \multicolumn{3}{c|}{Code Review}  \\ 
\cline{2-8}
                                                             & \#Refs & \#TCs &d-acc& r-acc & \#Refs & d-acc& r-acc                         \\ 
\hline
\begin{tabular}[c]{@{}l@{}}List-Compre\\\end{tabular} &  132                &   391                  & 1 & 1        & 100  & 1   & 1                              \\ 
\hline
\begin{tabular}[c]{@{}l@{}}Set-Compre\\\end{tabular}  & 21                &  39                     & 1   & 1      & 100  & 1     & 1                            \\ 
\hline
\begin{tabular}[c]{@{}l@{}}Dict-Compre\end{tabular}                                           & 102                 & 297                     & 1 & 1        & 100  & 1    & 1                             \\ 
\hline
\begin{tabular}[c]{@{}l@{}}Chain-Compa\end{tabular}                                               & 309                & 837                  & 1    & 1        & 100  & 1& 0.99                              \\ 
\hline
\begin{tabular}[c]{@{}l@{}}Truth-Test\end{tabular}                                               & 641                &  1680                  &  0.986   & 1   & 100  & 1   & 1                              \\ 
\hline
Loop-Else                                                    & 37                  &  98                 & 1    & 1        & 100  & 1   & 1                              \\ 
\hline
\begin{tabular}[c]{@{}l@{}}Assign-Multi-Tar\end{tabular}                                         & 1802               & 4565                & 0.999     &1     & 100  & 1  & 1                               \\ 
\hline
Star-in-Func-Call                                     & 86                &  201              & 0.977        &1     & 100  & 0.98      & 1                          \\ 
\hline
\begin{tabular}[c]{@{}l@{}}For-Multi-Tar\\\end{tabular}  & 85                 &  314                 & 1     & 1    & 100  & 1  & 1                               \\
\hline
\begin{tabular}[c]{@{}l@{}}Total\end{tabular}   & 3215                 &   7216               & 0.995     & 1    & 900  & 0.998  & 0.998 
                           
\\
\hline
\end{tabular}
\vspace{-0.5cm}
\end{table}
\vspace{-0.2cm}
\subsubsection{Result}

Table~\ref{tab:correctness} presents the analysis results.
\#Ref and \#TCs of the Testing column are the number of refactorings with successfully-executed test cases and the corresponding number of test cases. 
\#Ref of the Code Review column is the number of refactorings we reviewed. 
d-acc and r-acc are the accuracy of  non-idiomatic code smell detection and idiomatic code transformation respectively.
In total, we successfully test 3,215 refactorings from 479 repositories and reviewed 900 refactorings from 672 repositories.
Overall, our approach is very robust on real-world code.
It achieves 100\% accuracy of detection and refactoring for five pythonic idioms, i.e., list-comprehension, set-comprehension, dict-comprehension, loop-else and for-multi-targets.
It achieves 100\% detection accuracy for chain-comparison, and 100\% refactoring accuracy for truth-value-test, assign-multi-targets and star-in-func-call.

\textbf{Detection failure analysis.}
Our verification identifies 15 detected  non-idiomatic code fragments which are not refactorable, including 9 for truth-value-test, 2 for assign-multi-targets and 4 for star-in-func-call.
For example, for the truth-value-test \textsf{``if xpath\_results == []''}, if \textsf{``xpath\_results''} is an empty string, the if-condition is false. 
However, the idiom \textsf{``if not xpath\_results''} will be true if \textsf{``xpath\_results''} is an empty string.
Therefore, \textsf{``if xpath\_results == []''} cannot be refactored into \textsf{``if not xpath\_results''}. Other non-refactorable truth-value-test cases suffer from the same problem.

The two non-refactorable assign-multi-targets failures are
caused by the limitation of the Python static parsing. For example, for the two assignment statements, $stmt_1$: \textsf{lib=...} and $stmt_2$: \textsf{tmpl=`\{lib\}'}. 
\textsf{``lib''} of $stmt_2$ is a variable because Python uses curly brackets to insert variables in string. 
$stmt_2$ uses the target \textsf{``lib''} of $stmt_1$, so the two statements cannot be refactored into  \textsf{``lib,tmpl=...,`\{ lib\}'''} because it will make \textsf{tmpl} use the old value of \textsf{lib}. 
Since the parser parses \textsf{`\{lib\}'} into a string constant and does not parse the \textsf{``lib''} inside the brackets into a variable, we loss the information of data dependency and mistakenly identify the statements as refactorable.

For the star-in-func-call idiom, both testing and code review find two non-refactorable  non-idiomatic code fragments identified by our detection tool.
The reason is that our tool does not consider the semantic of Python slice. 
For example, for the code \textsf{``self.add\_circle\_arc(p[-1], p[0], p[1])''}, \textsf{``-1, 0, 1''} is an arithmetic sequence. However, Python list grows linearly and is not cyclic, as such slicing does not wrap (from end back to start going forward) as we expect.

\textbf{Code transformation failure analysis.}
Our tool makes only 1 code transformation error for chain-comparison.
For the code \textsf{``type is not None and self.\_meta\_types and type not in self.\_meta\_types''}, it has three comparison operations and \textsf{``type''} is the common operand of the first and the third operation. 
Therefore, we refactor it into \textsf{``None is not type not in self.\_meta\_types and self.\_
meta\_typestype''}. 
However, if \textsf{``self.\_meta\_types''} is None, the refactoring will report a TypeError at runtime because None is not iterable. 
As our tool does not analyze the priority of comparison operations and adjust their order when refactoring the code, the resulting code encounters the runtime error.

\noindent\fbox{\begin{minipage}{8.4cm} \emph{Our refactoring tool is robust and correct on real-world Python code. The limitation of Python static parsing and the complex program logic may result in a few rare detection and refactoring errors.} \end{minipage}}\\

\subsection{RQ2: Usefulness of Refactorings}

\subsubsection{Motivation}
Our tool is the first refactoring tool for pythonic idioms.
We would like to know how well Python developers accept the refactorings our tool makes and what opinions they have towards pythonic idiom refactoring in practice.

\subsubsection{Method}
We randomly sample 10 refactorings (including the original  non-idiomatic code fragments and the resulting idiomatic code after refactoring) for each type of pythonic idiom. 
The sampled refactorings come from 84 repositories.
We fork the repository corresponding to the  non-idiomatic code fragment and commit a pull request with the resulting idiomatic code. 
Readers can find the list of the 90 pull requests in \href{https://zenodo.org/record/6367738\#.YjRzLxBBzdo}{our replication package}. 
We collect the developers' responses to our pull requests, and count how many pull requests have been accepted or rejected by developers.
Among the accepted pull requests, we further count how many pull requests have been merged into the repositories.

\subsubsection{Result}

\begin{table}
\vspace{-0.4cm}
\small
\caption{Results of our refactoring pull requests}
\vspace{-0.4cm}
  \label{rq2_res}
\centering
\begin{tabular}{|l|c|c|c|c|} 
\hline
Idiom Category& Accepted & Rejected & Merged& \#Repo\\ 
\hline
List Comprehension      & 6   & 3 &5 &10\\ 
\hline
Set Comprehension       & 5    & 3 &4 &10   \\ 
\hline
Dict Comprehension     & 5    & 0 &5  &10\\ 
\hline
Chain Comparison      & 4    & 4 &3  &10\\ 
\hline
Truth Value Test       & 3    & 3 &3 &10\\ 
\hline
Loop Else              & 5    & 1 &4 &10\\ 
\hline
Assign Multi Targets    & 3    & 2 &1 &10\\ 
\hline
Star in Func Call & 1    & 6 &1  &10   \\ 
\hline
For Multi Targets       & 2    & 1 &2  &10\\
\hline
Total       & 34    & 23 &28  &84\\
\hline
\end{tabular}
\vspace{-0.3cm}
\end{table}

Table~\ref{rq2_res} presents our experiment results. 
Among 90 pull requests, we received 57 responses, including 34 accepted and 23 rejected.
28 of the accepted pull requests have been merged into the repositories.
The six pull requests that are not merged as they are not yet tested.
The 63\% (57/90) response rate indicates that Python developers pay attention to the pythonic idiom refactorings. 
The 60\% acceptance rate and the 50\% merge rate among the responses provide the initial evidence of our refactoring tool's practicality and usefulness.

Among the accepted pull requests, many developers praise the refactoring pull requests we made to their repositories. 
For example, two developers praise the dict-comprehension refactoring (\textsf{``\{
         name: mod for name, mod in module.named\_modules() if isinstance(mod, \_ConvNd)\}''}) ``\href{https://github.com/neuralmagic/sparseml/pull/510/files}{\textcolor{blue}{Thanks for the contribution, looks great!}}''. 
Many developers confirm that the suggested refactorings are more pythonic, such as \href{https://github.com/Serchinastico/Kin/pull/80}{``Definitely more pythonic!''} on a list-comprehension refactoring.
Some developers express the interests in refactoring other places with the same pythonic idiom, such as \href{https://github.com/AI4Finance-Foundation/ElegantRL/pull/101}{\textcolor{blue}{``I will change the other place to a more pythonic style ...''}} inspired by our pull request for an assign-multiple-targets refactoring. 

We analyze the rejected pull requests and summarize four main concerns developers have about pythonic idiom refactorings: \textit{readability}, \textit{performance}, \textit{systematic refactoring}, and \textit{inertia}.
  
\textbf{Readability.}  
   13 out of 23 reject responses are concerned that the pythonic idioms make the code less readable.
   For example, the developer comments on a suggested star-in-func-call refactoring ( \textsf{``*clip.size([2:4:1])''})
   : \href{https://github.com/facebookresearch/ClassyVision/pull/777}{\textcolor{blue}{``While your change is indeed feasible, I believe the original style is more readable''}}. 
   Even with the readability concern, some developers express that they learn something from the suggested refactorings. 
   For example, the developer comments on the chain-comparison refactoring (\textsf{``sessions is None is metrics''}):  \href{https://github.com/getsentry/sentry/pull/31245}{\textcolor{blue}{``... Interesting, ..., I learned something today though, thanks.''}}
   As another example, the developer worries that refactoring may loss specific information for the truth-value-test. 
   For example, a developer replied \href{https://github.com/mlrun/mlrun/pull/1660}{\textcolor{blue}{``I feel like asserting it to empty dict is more explicit and readable''}} if \textsf{``assert deepdiff.DeepDiff(...) == \{\}''} is refactored into \textsf{``assert not deepdiff.DeepDiff(...)''}.
     
\textbf{Performance.}
   3 reject responses are concerned about the performance or memory usage.
   For example, for the \href{https://github.com/google/jax/pull/9683}{\textcolor{blue}{list-comprehension refactoring}}, the developers reject the pull request because they are not sure that the performance improvement would be significant in their project.
   For the \href{https://github.com/micropython/micropython-lib/pull/477}{\textcolor{blue}{star-in-func-call refactoring}} which refactors \textsf{``ss[0], ss[1], ss[2], ss[3], ss[4], ss[5], ss[6]''} into \textsf{``*ss[0:7:1]''}, the developer believes the refactoring can cause memory fragmentation.
  
   
\textbf{Systematic refactoring.}
    3 reject responses indicates that developers do not want to refactor the project in an ad-hoc way.
    Two responses are discouraged to refactor only one code fragment of the project. For example, although the developers reject \href{https://github.com/streamlink/streamlink/pull/4303}{\textcolor{blue}{our pull request for a set-comprehension refactoring}}, they propose that such refactorings should be applied to the whole project rather than by a single pull request to just one place.
    In another reject response, the developer replies that \href{https://github.com/projecthamster/hamster/pull/696}{\textcolor{blue}{``Waf is just a tool for us. We don't need style patches for it.''}} for a list-comprehension refactoring.
    In fact, we believe these responses confirm the need for systematic pythonic idiom refactoring tool like ours.
    Our tool can scan and refactor the whole project and dependent packages. 
    It was just we submitted only some randomly sampled refactorings to the projects.    
 
\textbf{Inertia.} 
4 rejects are because the developers prefer the original code. For example, a developer replies ``\href{https://github.com/apache/tvm/pull/10002}{\textcolor{blue}{Thanks for the suggestion, I prefer the existing code}}'' for a star-in-func-call refactoring.
And some developers would like to accept pull requests to fix bugs instead of code refactoring, e.g., the developer replies to a for-multiple-targets refactoring: \href{https://github.com/sympy/sympy/pull/23065}{\textcolor{blue}{``I think it's better to leave the RUBI code alone for now unless there is work to fix it.''}}.

\vspace{1mm}
\noindent\fbox{\begin{minipage}{8.4cm} \emph{Our pythonic idiom refactorings have been well received by the developers. The developers also raise concerns about readability and performance of pythonic idioms which deserve further study.}
\end{minipage}}\\

%% file: related.tex
\subsection{Pythonic Idioms}
How to write pythonic code has been a popular topic~\cite{alexandru2018usage,knupp2013writing,hettinger2013transforming,slatkin2019effective,merchantepython}.
Several studies identify some Python idioms to help developers write more idiomatic code (commonly referred to as more pythonic~\cite{van2007python,merchantepython}). 
Alexandru et al.~\cite{alexandru2018usage} identify 19 pythonic idioms (list-comprehension and dict-comprehension overlap with our work) from several books.
Merchante et al.~\cite{merchantepython} identify the usage of pythonic idioms in GitHub. 
Kula et al.~\cite{sakulniwat2019visualizing} visualize the usage of the with-open idiom over time, and find that projects are accustomed to using the with-open idiom instead of non-idiomatic counterpart. 
Phan-udom et al.~\cite{phan2020teddy} develop a tool to recommend pythonic code examples by searching similar code examples from 113 code snippets of three repositories. 
Different from these related works, our work identifies unique pythonic idioms from the language syntax and find 4 idioms (star-in-func-call, for-multi-targets, assign-multi-targets, loop-else) that have not been identified before. 
Furthermore, we not only identify pythonic idioms but also develop an automatic tool to refactor anti-idiom code into idiomatic code. 

\subsection{Code refactoring}
Martin Fowler proposes code refactoring~\cite{fowler2018refactoring} about 30 years ago. An active line of research is to recover refactorings made in the code~\cite{xing2006refactoring,dig2006automated,godfrey2005using,kim2010ref,negara2013comparative, tsantalis2018accurate,silva2017refdiff,weissgerber2006identifying, demeyer2000finding,prete2010template}. 
Prete et al.~\cite{prete2010template,kim2010ref} detect the largest number of refactoring types based on the Fowler’s catalog. 
Tsantalis et. al.~\cite{tsantalis2018accurate} identify 15 Java refactoring types by statement matching algorithm and refactoring detection rules. 
Dilhara et. al.~\cite{dilhara2022discovering} identify 18 Python refactoring kinds (e.g., rename, move, pull up methods), which do not overlap with our refactoring type. 
Another active line of research is to refactor code~\cite{franklin2013lambdaficator,radoi2014translating,ouni2016multi,tsantalis2017clone,david2017kayak,kohler2019automated,midolo2021refactoring,Pylint}. 
Ouni et al.~\cite{ouni2016multi} propose a multi-objective search-based approach for finding the optimal sequence of refactorings. 
K{\"o}hler et al.~\cite{kohler2019automated} develop a tool to automatically convert asynchronous code to reactive programming. 
Pylint ~\cite{Pylint} is a static code analysis tool, which could give refactoring suggestions for chain-comparison and truth-value-test, but it does not support automatic refactorings. 
To the best of our knowledge, our tool is the first automatic refactoring tool which covers 9 types of pythonic idioms.

%% file: main_arkiv.bbl

\begin{thebibliography}{65}


\ifx \showCODEN    \undefined \def \showCODEN     #1{\unskip}     \fi
\ifx \showDOI      \undefined \def \showDOI       #1{#1}\fi
\ifx \showISBNx    \undefined \def \showISBNx     #1{\unskip}     \fi
\ifx \showISBNxiii \undefined \def \showISBNxiii  #1{\unskip}     \fi
\ifx \showISSN     \undefined \def \showISSN      #1{\unskip}     \fi
\ifx \showLCCN     \undefined \def \showLCCN      #1{\unskip}     \fi
\ifx \shownote     \undefined \def \shownote      #1{#1}          \fi
\ifx \showarticletitle \undefined \def \showarticletitle #1{#1}   \fi
\ifx \showURL      \undefined \def \showURL       {\relax}        \fi
\providecommand\bibfield[2]{#2}
\providecommand\bibinfo[2]{#2}
\providecommand\natexlab[1]{#1}
\providecommand\showeprint[2][]{arXiv:#2}

\bibitem[\protect\citeauthoryear{??}{Pyl}{2021}]%
        {Pylint}
 \bibinfo{year}{2021}\natexlab{}.
\newblock \bibinfo{booktitle}{\emph{Pylint}}.
\newblock
\urldef\tempurl%
\url{https://pylint.org/}
\showURL{%
\tempurl}


\bibitem[\protect\citeauthoryear{??}{mul}{2022a}]%
        {multi_assign_loginradius}
 \bibinfo{year}{2022}\natexlab{a}.
\newblock \bibinfo{booktitle}{\emph{Assign with Multiple Targets on
  loginradius}}.
\newblock
\urldef\tempurl%
\url{https://www.loginradius.com/blog/async/speed-up-python-code/}
\showURL{%
\tempurl}


\bibitem[\protect\citeauthoryear{??}{mul}{2022b}]%
        {multi_assign_medium}
 \bibinfo{year}{2022}\natexlab{b}.
\newblock \bibinfo{booktitle}{\emph{Assign with Multiple Targets on Medium}}.
\newblock
\urldef\tempurl%
\url{https://medium.com/geekculture/3-easy-ways-to-instantly-make-your-python-program-faster-e599e920ea28}
\showURL{%
\tempurl}


\bibitem[\protect\citeauthoryear{??}{mul}{2022c}]%
        {multi_assign_so}
 \bibinfo{year}{2022}\natexlab{c}.
\newblock \bibinfo{booktitle}{\emph{Assign with Multiple Targets on
  StackOverflow}}.
\newblock
\urldef\tempurl%
\url{https://stackoverflow.com/questions/22278695/python-multiple-assignment-vs-individual-assignment-speed}
\showURL{%
\tempurl}


\bibitem[\protect\citeauthoryear{??}{cha}{2022a}]%
        {chainCompare_so}
 \bibinfo{year}{2022}\natexlab{a}.
\newblock \bibinfo{booktitle}{\emph{Chain Comparison on StackOverflow}}.
\newblock
\urldef\tempurl%
\url{https://stackoverflow.com/questions/48375753/why-are-chained-operator-expressions-slower-than-their-expanded-equivalent}
\showURL{%
\tempurl}


\bibitem[\protect\citeauthoryear{??}{cha}{2022b}]%
        {chainCompare_perfor_wiki}
 \bibinfo{year}{2022}\natexlab{b}.
\newblock \bibinfo{booktitle}{\emph{Chain Comparison on Wiki}}.
\newblock
\urldef\tempurl%
\url{https://wiki.python.org/moin/PythonSpeed}
\showURL{%
\tempurl}


\bibitem[\protect\citeauthoryear{??}{dic}{2022a}]%
        {dict_comprehension_performance}
 \bibinfo{year}{2022}\natexlab{a}.
\newblock \bibinfo{booktitle}{\emph{Dict Comprehension}}.
\newblock
\urldef\tempurl%
\url{https://stackoverflow.com/questions/52542742/why-is-this-loop-faster-than-a-dictionary-comprehension-for-creating-a-dictionar}
\showURL{%
\tempurl}


\bibitem[\protect\citeauthoryear{??}{for}{2022a}]%
        {for_targets_so2}
 \bibinfo{year}{2022}\natexlab{a}.
\newblock \bibinfo{booktitle}{\emph{For Multiple Targets on StackOverflow}}.
\newblock
\urldef\tempurl%
\url{https://stackoverflow.com/questions/13024416/how-come-unpacking-is-faster-than-accessing-by-index}
\showURL{%
\tempurl}


\bibitem[\protect\citeauthoryear{??}{for}{2022b}]%
        {for_targets_so1}
 \bibinfo{year}{2022}\natexlab{b}.
\newblock \bibinfo{booktitle}{\emph{For Multiple Targets on StackOverflow}}.
\newblock
\urldef\tempurl%
\url{https://stackoverflow.com/questions/23039485/for-loop-item-unpacking}
\showURL{%
\tempurl}


\bibitem[\protect\citeauthoryear{??}{lis}{2022a}]%
        {list_compr_10X_faster}
 \bibinfo{year}{2022}\natexlab{a}.
\newblock \bibinfo{booktitle}{\emph{List comprehension is 10X faster than
  loops}}.
\newblock
\urldef\tempurl%
\url{https://innovationyourself.com/list-comprehension-in-python/}
\showURL{%
\tempurl}


\bibitem[\protect\citeauthoryear{??}{lis}{2022b}]%
        {list_comprehension_performance}
 \bibinfo{year}{2022}\natexlab{b}.
\newblock \bibinfo{booktitle}{\emph{List Comprehension on StackOverflow}}.
\newblock
\urldef\tempurl%
\url{https://stackoverflow.com/questions/30245397/why-is-a-list-comprehension-so-much-faster-than-appending-to-a-list}
\showURL{%
\tempurl}


\bibitem[\protect\citeauthoryear{??}{loo}{2022}]%
        {loop_else_performance_so}
 \bibinfo{year}{2022}\natexlab{}.
\newblock \bibinfo{booktitle}{\emph{Loop Else}}.
\newblock
\urldef\tempurl%
\url{https://stackoverflow.com/questions/13069402/efficient-implementation-for-python-for-else-loop-in-java}
\showURL{%
\tempurl}


\bibitem[\protect\citeauthoryear{??}{Med}{2022}]%
        {Medium}
 \bibinfo{year}{2022}\natexlab{}.
\newblock \bibinfo{booktitle}{\emph{Medium}}.
\newblock
\urldef\tempurl%
\url{https://medium.com/}
\showURL{%
\tempurl}


\bibitem[\protect\citeauthoryear{??}{lis}{2022c}]%
        {list_compr_pep202}
 \bibinfo{year}{2022}\natexlab{c}.
\newblock \bibinfo{booktitle}{\emph{PEP 202-List comprehension}}.
\newblock
\urldef\tempurl%
\url{https://peps.python.org/pep-0202/}
\showURL{%
\tempurl}


\bibitem[\protect\citeauthoryear{??}{dic}{2022b}]%
        {dict_compr_pep274}
 \bibinfo{year}{2022}\natexlab{b}.
\newblock \bibinfo{booktitle}{\emph{PEP 274-Dict comprehension}}.
\newblock
\urldef\tempurl%
\url{https://peps.python.org/pep-0274/}
\showURL{%
\tempurl}


\bibitem[\protect\citeauthoryear{??}{sta}{2022}]%
        {star_pep_448}
 \bibinfo{year}{2022}\natexlab{}.
\newblock \bibinfo{booktitle}{\emph{PEP 448-Additional Unpacking
  Generalizations}}.
\newblock
\urldef\tempurl%
\url{https://peps.python.org/pep-0202/}
\showURL{%
\tempurl}


\bibitem[\protect\citeauthoryear{??}{pro}{2022}]%
        {programming_idioms}
 \bibinfo{year}{2022}\natexlab{}.
\newblock \bibinfo{booktitle}{\emph{Programming Idioms}}.
\newblock
\urldef\tempurl%
\url{https://programming-idioms.org/}
\showURL{%
\tempurl}


\bibitem[\protect\citeauthoryear{??}{pyt}{2022}]%
        {pythonAPI}
 \bibinfo{year}{2022}\natexlab{}.
\newblock \bibinfo{booktitle}{\emph{Python Abstract Grammar}}.
\newblock
\urldef\tempurl%
\url{https://docs.python.org/3/library/ast.html}
\showURL{%
\tempurl}


\bibitem[\protect\citeauthoryear{??}{cha}{2022c}]%
        {chainCompare}
 \bibinfo{year}{2022}\natexlab{c}.
\newblock \bibinfo{booktitle}{\emph{Python Chain Comparison}}.
\newblock
\urldef\tempurl%
\url{https://docs.python.org/3/reference/expressions.html#comparisons}
\showURL{%
\tempurl}


\bibitem[\protect\citeauthoryear{??}{set}{2022}]%
        {set_comprehension_performance}
 \bibinfo{year}{2022}\natexlab{}.
\newblock \bibinfo{booktitle}{\emph{Set Comprehension}}.
\newblock
\urldef\tempurl%
\url{https://appdividend.com/2020/12/03/python-set-comprehension/}
\showURL{%
\tempurl}


\bibitem[\protect\citeauthoryear{??}{Sta}{2022}]%
        {StackOverflow}
 \bibinfo{year}{2022}\natexlab{}.
\newblock \bibinfo{booktitle}{\emph{StackOverflow}}.
\newblock
\urldef\tempurl%
\url{https://stackoverflow.com/}
\showURL{%
\tempurl}


\bibitem[\protect\citeauthoryear{??}{sta}{2022}]%
        {star_so}
 \bibinfo{year}{2022}\natexlab{}.
\newblock \bibinfo{booktitle}{\emph{Star in Call on StackOverflow}}.
\newblock
\urldef\tempurl%
\url{https://stackoverflow.com/questions/2921847/what-does-the-star-and-doublestar-operator-mean-in-a-function-call}
\showURL{%
\tempurl}


\bibitem[\protect\citeauthoryear{??}{tim}{2022}]%
        {timeit}
 \bibinfo{year}{2022}\natexlab{}.
\newblock \bibinfo{booktitle}{\emph{timeit}}.
\newblock
\urldef\tempurl%
\url{https://docs.python.org/3/library/timeit.html}
\showURL{%
\tempurl}


\bibitem[\protect\citeauthoryear{??}{tru}{2022a}]%
        {truth_so}
 \bibinfo{year}{2022}\natexlab{a}.
\newblock \bibinfo{booktitle}{\emph{Truth Value Test}}.
\newblock
\urldef\tempurl%
\url{https://stackoverflow.com/questions/39983695/what-is-truthy-and-falsy-how-is-it-different-from-true-and-false}
\showURL{%
\tempurl}


\bibitem[\protect\citeauthoryear{??}{lis}{2022d}]%
        {list_truth_so}
 \bibinfo{year}{2022}\natexlab{d}.
\newblock \bibinfo{booktitle}{\emph{Truth Value Test}}.
\newblock
\urldef\tempurl%
\url{https://stackoverflow.com/questions/53513/how-do-i-check-if-a-list-is-empty}
\showURL{%
\tempurl}


\bibitem[\protect\citeauthoryear{??}{tru}{2022b}]%
        {truth_value_test_python_doc}
 \bibinfo{year}{2022}\natexlab{b}.
\newblock \bibinfo{booktitle}{\emph{Truth Value Test on Python Documentation}}.
\newblock
\urldef\tempurl%
\url{https://docs.python.org/3/library/stdtypes.html#truth-value-testing}
\showURL{%
\tempurl}


\bibitem[\protect\citeauthoryear{??}{uni}{2022}]%
        {unittest}
 \bibinfo{year}{2022}\natexlab{}.
\newblock \bibinfo{booktitle}{\emph{Unit testing framework}}.
\newblock
\urldef\tempurl%
\url{https://docs.python.org/3/library/unittest.html}
\showURL{%
\tempurl}


\bibitem[\protect\citeauthoryear{??}{wri}{2022}]%
        {writing_fast_python}
 \bibinfo{year}{2022}\natexlab{}.
\newblock \bibinfo{booktitle}{\emph{Writing Fast Python}}.
\newblock
\urldef\tempurl%
\url{https://switowski.com/blog/checking-for-true-or-false}
\showURL{%
\tempurl}


\bibitem[\protect\citeauthoryear{Alexandru, Merchante, Panichella, Proksch,
  Gall, and Robles}{Alexandru et~al\mbox{.}}{2018}]%
        {alexandru2018usage}
\bibfield{author}{\bibinfo{person}{Carol~V Alexandru},
  \bibinfo{person}{Jos{\'e}~J Merchante}, \bibinfo{person}{Sebastiano
  Panichella}, \bibinfo{person}{Sebastian Proksch}, \bibinfo{person}{Harald~C
  Gall}, {and} \bibinfo{person}{Gregorio Robles}.}
  \bibinfo{year}{2018}\natexlab{}.
\newblock \showarticletitle{On the usage of pythonic idioms}. In
  \bibinfo{booktitle}{\emph{Proceedings of the 2018 ACM SIGPLAN International
  Symposium on New Ideas, New Paradigms, and Reflections on Programming and
  Software}}. \bibinfo{pages}{1--11}.
\newblock


\bibitem[\protect\citeauthoryear{Allamanis, Barr, Bird, Devanbu, Marron, and
  Sutton}{Allamanis et~al\mbox{.}}{2018}]%
        {allamanis2018mining}
\bibfield{author}{\bibinfo{person}{Miltiadis Allamanis},
  \bibinfo{person}{Earl~T Barr}, \bibinfo{person}{Christian Bird},
  \bibinfo{person}{Premkumar Devanbu}, \bibinfo{person}{Mark Marron}, {and}
  \bibinfo{person}{Charles Sutton}.} \bibinfo{year}{2018}\natexlab{}.
\newblock \showarticletitle{Mining semantic loop idioms}.
\newblock \bibinfo{journal}{\emph{IEEE Transactions on Software Engineering}}
  \bibinfo{volume}{44}, \bibinfo{number}{7} (\bibinfo{year}{2018}),
  \bibinfo{pages}{651--668}.
\newblock


\bibitem[\protect\citeauthoryear{Allamanis and Sutton}{Allamanis and
  Sutton}{2014}]%
        {allamanis2014mining}
\bibfield{author}{\bibinfo{person}{Miltiadis Allamanis} {and}
  \bibinfo{person}{Charles Sutton}.} \bibinfo{year}{2014}\natexlab{}.
\newblock \showarticletitle{Mining idioms from source code}. In
  \bibinfo{booktitle}{\emph{Proceedings of the 22nd acm sigsoft international
  symposium on foundations of software engineering}}.
  \bibinfo{pages}{472--483}.
\newblock


\bibitem[\protect\citeauthoryear{David, Kesseli, and Kroening}{David
  et~al\mbox{.}}{2017}]%
        {david2017kayak}
\bibfield{author}{\bibinfo{person}{Cristina David}, \bibinfo{person}{Pascal
  Kesseli}, {and} \bibinfo{person}{Daniel Kroening}.}
  \bibinfo{year}{2017}\natexlab{}.
\newblock \showarticletitle{Kayak: Safe semantic refactoring to java streams}.
\newblock \bibinfo{journal}{\emph{arXiv preprint arXiv:1712.07388}}
  (\bibinfo{year}{2017}).
\newblock


\bibitem[\protect\citeauthoryear{Demeyer, Ducasse, and Nierstrasz}{Demeyer
  et~al\mbox{.}}{2000}]%
        {demeyer2000finding}
\bibfield{author}{\bibinfo{person}{Serge Demeyer},
  \bibinfo{person}{St{\'e}phane Ducasse}, {and} \bibinfo{person}{Oscar
  Nierstrasz}.} \bibinfo{year}{2000}\natexlab{}.
\newblock \showarticletitle{Finding refactorings via change metrics}.
\newblock \bibinfo{journal}{\emph{ACM SIGPLAN Notices}} \bibinfo{volume}{35},
  \bibinfo{number}{10} (\bibinfo{year}{2000}), \bibinfo{pages}{166--177}.
\newblock


\bibitem[\protect\citeauthoryear{Dig, Comertoglu, Marinov, and Johnson}{Dig
  et~al\mbox{.}}{2006}]%
        {dig2006automated}
\bibfield{author}{\bibinfo{person}{Danny Dig}, \bibinfo{person}{Can
  Comertoglu}, \bibinfo{person}{Darko Marinov}, {and} \bibinfo{person}{Ralph
  Johnson}.} \bibinfo{year}{2006}\natexlab{}.
\newblock \showarticletitle{Automated detection of refactorings in evolving
  components}. In \bibinfo{booktitle}{\emph{European conference on
  object-oriented programming}}. Springer, \bibinfo{pages}{404--428}.
\newblock


\bibitem[\protect\citeauthoryear{Dilhara, Ketkar, Sannidhi, and Dig}{Dilhara
  et~al\mbox{.}}{2022}]%
        {dilhara2022discovering}
\bibfield{author}{\bibinfo{person}{Malinda Dilhara}, \bibinfo{person}{Ameya
  Ketkar}, \bibinfo{person}{Nikhith Sannidhi}, {and} \bibinfo{person}{Danny
  Dig}.} \bibinfo{year}{2022}\natexlab{}.
\newblock \showarticletitle{Discovering Repetitive Code Changes in Python ML
  Systems}. In \bibinfo{booktitle}{\emph{International Conference on Software
  Engineering (ICSE’22). To appear}}.
\newblock


\bibitem[\protect\citeauthoryear{Farooq and Zaytsev}{Farooq and
  Zaytsev}{2021}]%
        {Zen_Your_Python}
\bibfield{author}{\bibinfo{person}{Aamir Farooq} {and} \bibinfo{person}{Vadim
  Zaytsev}.} \bibinfo{year}{2021}\natexlab{}.
\newblock \showarticletitle{There is More than One Way to Zen Your Python}. In
  \bibinfo{booktitle}{\emph{Proceedings of the 14th ACM SIGPLAN International
  Conference on Software Language Engineering}}. \bibinfo{pages}{68–82}.
\newblock


\bibitem[\protect\citeauthoryear{Fowler}{Fowler}{2018}]%
        {fowler2018refactoring}
\bibfield{author}{\bibinfo{person}{Martin Fowler}.}
  \bibinfo{year}{2018}\natexlab{}.
\newblock \bibinfo{booktitle}{\emph{Refactoring: improving the design of
  existing code}}.
\newblock \bibinfo{publisher}{Addison-Wesley Professional}.
\newblock


\bibitem[\protect\citeauthoryear{Franklin, Gyori, Lahoda, and Dig}{Franklin
  et~al\mbox{.}}{2013}]%
        {franklin2013lambdaficator}
\bibfield{author}{\bibinfo{person}{Lyle Franklin}, \bibinfo{person}{Alex
  Gyori}, \bibinfo{person}{Jan Lahoda}, {and} \bibinfo{person}{Danny Dig}.}
  \bibinfo{year}{2013}\natexlab{}.
\newblock \showarticletitle{LAMBDAFICATOR: from imperative to functional
  programming through automated refactoring}. In \bibinfo{booktitle}{\emph{2013
  35th international conference on software engineering (ICSE)}}. IEEE,
  \bibinfo{pages}{1287--1290}.
\newblock


\bibitem[\protect\citeauthoryear{Gao, Xia, Lo, Grundy, and Zimmermann}{Gao
  et~al\mbox{.}}{2021}]%
        {gao2021automating}
\bibfield{author}{\bibinfo{person}{Zhipeng Gao}, \bibinfo{person}{Xin Xia},
  \bibinfo{person}{David Lo}, \bibinfo{person}{John Grundy}, {and}
  \bibinfo{person}{Thomas Zimmermann}.} \bibinfo{year}{2021}\natexlab{}.
\newblock \showarticletitle{Automating the removal of obsolete TODO comments}.
  In \bibinfo{booktitle}{\emph{Proceedings of the 29th ACM Joint Meeting on
  European Software Engineering Conference and Symposium on the Foundations of
  Software Engineering}}. \bibinfo{pages}{218--229}.
\newblock


\bibitem[\protect\citeauthoryear{Godfrey and Zou}{Godfrey and Zou}{2005}]%
        {godfrey2005using}
\bibfield{author}{\bibinfo{person}{Michael~W Godfrey} {and}
  \bibinfo{person}{Lijie Zou}.} \bibinfo{year}{2005}\natexlab{}.
\newblock \showarticletitle{Using origin analysis to detect merging and
  splitting of source code entities}.
\newblock \bibinfo{journal}{\emph{IEEE Transactions on Software Engineering}}
  \bibinfo{volume}{31}, \bibinfo{number}{2} (\bibinfo{year}{2005}),
  \bibinfo{pages}{166--181}.
\newblock


\bibitem[\protect\citeauthoryear{Hettinger}{Hettinger}{2013}]%
        {hettinger2013transforming}
\bibfield{author}{\bibinfo{person}{Raymond Hettinger}.}
  \bibinfo{year}{2013}\natexlab{}.
\newblock \bibinfo{booktitle}{\emph{Transforming code into beautiful, idiomatic
  Python}}.
\newblock
\urldef\tempurl%
\url{https://www.youtube.com/watch?v=OSGv2VnC0go}
\showURL{%
\tempurl}


\bibitem[\protect\citeauthoryear{Hunt}{Hunt}{2019}]%
        {hunt2019pytest}
\bibfield{author}{\bibinfo{person}{John Hunt}.}
  \bibinfo{year}{2019}\natexlab{}.
\newblock \showarticletitle{PyTest Testing Framework}.
\newblock In \bibinfo{booktitle}{\emph{Advanced Guide to Python 3
  Programming}}. \bibinfo{publisher}{Springer}, \bibinfo{pages}{175--186}.
\newblock


\bibitem[\protect\citeauthoryear{Kim, Gee, Loh, and Rachatasumrit}{Kim
  et~al\mbox{.}}{2010}]%
        {kim2010ref}
\bibfield{author}{\bibinfo{person}{Miryung Kim}, \bibinfo{person}{Matthew Gee},
  \bibinfo{person}{Alex Loh}, {and} \bibinfo{person}{Napol Rachatasumrit}.}
  \bibinfo{year}{2010}\natexlab{}.
\newblock \showarticletitle{Ref-finder: a refactoring reconstruction tool based
  on logic query templates}. In \bibinfo{booktitle}{\emph{Proceedings of the
  eighteenth ACM SIGSOFT international symposium on Foundations of software
  engineering}}. \bibinfo{pages}{371--372}.
\newblock


\bibitem[\protect\citeauthoryear{Knupp}{Knupp}{2013}]%
        {knupp2013writing}
\bibfield{author}{\bibinfo{person}{Jeff Knupp}.}
  \bibinfo{year}{2013}\natexlab{}.
\newblock \bibinfo{booktitle}{\emph{Writing Idiomatic Python 3.3}}.
\newblock \bibinfo{publisher}{Jeff Knupp}.
\newblock


\bibitem[\protect\citeauthoryear{K{\"o}hler and Salvaneschi}{K{\"o}hler and
  Salvaneschi}{2019}]%
        {kohler2019automated}
\bibfield{author}{\bibinfo{person}{Mirko K{\"o}hler} {and}
  \bibinfo{person}{Guido Salvaneschi}.} \bibinfo{year}{2019}\natexlab{}.
\newblock \showarticletitle{Automated refactoring to reactive programming}. In
  \bibinfo{booktitle}{\emph{2019 34th IEEE/ACM International Conference on
  Automated Software Engineering (ASE)}}. IEEE, \bibinfo{pages}{835--846}.
\newblock


\bibitem[\protect\citeauthoryear{Merchante and Robles}{Merchante and
  Robles}{2017}]%
        {merchantepython}
\bibfield{author}{\bibinfo{person}{Jos{\'e}~Javier Merchante} {and}
  \bibinfo{person}{Gregorio Robles}.} \bibinfo{year}{2017}\natexlab{}.
\newblock \showarticletitle{From Python to Pythonic: Searching for Python
  idioms in GitHub}. In \bibinfo{booktitle}{\emph{Proceedings of the Seminar
  Series on Advanced Techniques and Tools for Software Evolution}}.
  \bibinfo{pages}{1--3}.
\newblock


\bibitem[\protect\citeauthoryear{Midolo and Tramontana}{Midolo and
  Tramontana}{2021}]%
        {midolo2021refactoring}
\bibfield{author}{\bibinfo{person}{Alessandro Midolo} {and}
  \bibinfo{person}{Emiliano Tramontana}.} \bibinfo{year}{2021}\natexlab{}.
\newblock \showarticletitle{Refactoring Java Loops to Streams Automatically}.
  In \bibinfo{booktitle}{\emph{2021 4th International Conference on Computer
  Science and Software Engineering (CSSE 2021)}}. \bibinfo{pages}{135--139}.
\newblock


\bibitem[\protect\citeauthoryear{Negara, Chen, Vakilian, Johnson, and
  Dig}{Negara et~al\mbox{.}}{2013}]%
        {negara2013comparative}
\bibfield{author}{\bibinfo{person}{Stas Negara}, \bibinfo{person}{Nicholas
  Chen}, \bibinfo{person}{Mohsen Vakilian}, \bibinfo{person}{Ralph~E Johnson},
  {and} \bibinfo{person}{Danny Dig}.} \bibinfo{year}{2013}\natexlab{}.
\newblock \showarticletitle{A comparative study of manual and automated
  refactorings}. In \bibinfo{booktitle}{\emph{European Conference on
  Object-Oriented Programming}}. Springer, \bibinfo{pages}{552--576}.
\newblock


\bibitem[\protect\citeauthoryear{Ouni, Kessentini, Sahraoui, Inoue, and
  Deb}{Ouni et~al\mbox{.}}{2016}]%
        {ouni2016multi}
\bibfield{author}{\bibinfo{person}{Ali Ouni}, \bibinfo{person}{Marouane
  Kessentini}, \bibinfo{person}{Houari Sahraoui}, \bibinfo{person}{Katsuro
  Inoue}, {and} \bibinfo{person}{Kalyanmoy Deb}.}
  \bibinfo{year}{2016}\natexlab{}.
\newblock \showarticletitle{Multi-criteria code refactoring using search-based
  software engineering: An industrial case study}.
\newblock \bibinfo{journal}{\emph{ACM Transactions on Software Engineering and
  Methodology (TOSEM)}} \bibinfo{volume}{25}, \bibinfo{number}{3}
  (\bibinfo{year}{2016}), \bibinfo{pages}{1--53}.
\newblock


\bibitem[\protect\citeauthoryear{Pan, Huang, Wang, Zhang, and Li}{Pan
  et~al\mbox{.}}{2020}]%
        {pan2020reinforcement}
\bibfield{author}{\bibinfo{person}{Minxue Pan}, \bibinfo{person}{An Huang},
  \bibinfo{person}{Guoxin Wang}, \bibinfo{person}{Tian Zhang}, {and}
  \bibinfo{person}{Xuandong Li}.} \bibinfo{year}{2020}\natexlab{}.
\newblock \showarticletitle{Reinforcement learning based curiosity-driven
  testing of Android applications}. In \bibinfo{booktitle}{\emph{Proceedings of
  the 29th ACM SIGSOFT International Symposium on Software Testing and
  Analysis}}. \bibinfo{pages}{153--164}.
\newblock


\bibitem[\protect\citeauthoryear{Phan-udom, Wattanakul, Sakulniwat,
  Ragkhitwetsagul, Sunetnanta, Choetkiertikul, and Kula}{Phan-udom
  et~al\mbox{.}}{2020}]%
        {phan2020teddy}
\bibfield{author}{\bibinfo{person}{Purit Phan-udom}, \bibinfo{person}{Naruedon
  Wattanakul}, \bibinfo{person}{Tattiya Sakulniwat}, \bibinfo{person}{Chaiyong
  Ragkhitwetsagul}, \bibinfo{person}{Thanwadee Sunetnanta},
  \bibinfo{person}{Morakot Choetkiertikul}, {and}
  \bibinfo{person}{Raula~Gaikovina Kula}.} \bibinfo{year}{2020}\natexlab{}.
\newblock \showarticletitle{Teddy: Automatic Recommendation of Pythonic Idiom
  Usage For Pull-Based Software Projects}. In \bibinfo{booktitle}{\emph{2020
  IEEE International Conference on Software Maintenance and Evolution
  (ICSME)}}. IEEE, \bibinfo{pages}{806--809}.
\newblock


\bibitem[\protect\citeauthoryear{Prete, Rachatasumrit, Sudan, and Kim}{Prete
  et~al\mbox{.}}{2010}]%
        {prete2010template}
\bibfield{author}{\bibinfo{person}{Kyle Prete}, \bibinfo{person}{Napol
  Rachatasumrit}, \bibinfo{person}{Nikita Sudan}, {and}
  \bibinfo{person}{Miryung Kim}.} \bibinfo{year}{2010}\natexlab{}.
\newblock \showarticletitle{Template-based reconstruction of complex
  refactorings}. In \bibinfo{booktitle}{\emph{2010 IEEE International
  Conference on Software Maintenance}}. IEEE, \bibinfo{pages}{1--10}.
\newblock


\bibitem[\protect\citeauthoryear{Radoi, Fink, Rabbah, and Sridharan}{Radoi
  et~al\mbox{.}}{2014}]%
        {radoi2014translating}
\bibfield{author}{\bibinfo{person}{Cosmin Radoi}, \bibinfo{person}{Stephen~J
  Fink}, \bibinfo{person}{Rodric Rabbah}, {and} \bibinfo{person}{Manu
  Sridharan}.} \bibinfo{year}{2014}\natexlab{}.
\newblock \showarticletitle{Translating imperative code to MapReduce}. In
  \bibinfo{booktitle}{\emph{Proceedings of the 2014 ACM International
  Conference on Object Oriented Programming Systems Languages \&
  Applications}}. \bibinfo{pages}{909--927}.
\newblock


\bibitem[\protect\citeauthoryear{Reitz and Schlusser}{Reitz and
  Schlusser}{2016}]%
        {reitz2016hitchhiker}
\bibfield{author}{\bibinfo{person}{Kenneth Reitz} {and} \bibinfo{person}{Tanya
  Schlusser}.} \bibinfo{year}{2016}\natexlab{}.
\newblock \bibinfo{booktitle}{\emph{The Hitchhiker's guide to Python: best
  practices for development}}.
\newblock \bibinfo{publisher}{" O'Reilly Media, Inc."}.
\newblock


\bibitem[\protect\citeauthoryear{Sakulniwat, Kula, Ragkhitwetsagul,
  Choetkiertikul, Sunetnanta, Wang, Ishio, and Matsumoto}{Sakulniwat
  et~al\mbox{.}}{2019}]%
        {sakulniwat2019visualizing}
\bibfield{author}{\bibinfo{person}{Tattiya Sakulniwat},
  \bibinfo{person}{Raula~Gaikovina Kula}, \bibinfo{person}{Chaiyong
  Ragkhitwetsagul}, \bibinfo{person}{Morakot Choetkiertikul},
  \bibinfo{person}{Thanwadee Sunetnanta}, \bibinfo{person}{Dong Wang},
  \bibinfo{person}{Takashi Ishio}, {and} \bibinfo{person}{Kenichi Matsumoto}.}
  \bibinfo{year}{2019}\natexlab{}.
\newblock \showarticletitle{Visualizing the Usage of Pythonic Idioms Over Time:
  A Case Study of the with open Idiom}. In \bibinfo{booktitle}{\emph{2019 10th
  International Workshop on Empirical Software Engineering in Practice
  (IWESEP)}}. IEEE, \bibinfo{pages}{43--435}.
\newblock


\bibitem[\protect\citeauthoryear{Silva and Valente}{Silva and Valente}{2017}]%
        {silva2017refdiff}
\bibfield{author}{\bibinfo{person}{Danilo Silva} {and}
  \bibinfo{person}{Marco~Tulio Valente}.} \bibinfo{year}{2017}\natexlab{}.
\newblock \showarticletitle{Refdiff: detecting refactorings in version
  histories}. In \bibinfo{booktitle}{\emph{2017 IEEE/ACM 14th International
  Conference on Mining Software Repositories (MSR)}}. IEEE,
  \bibinfo{pages}{269--279}.
\newblock


\bibitem[\protect\citeauthoryear{Sivaraman, Abreu, Scott, Akomolede, and
  Chandra}{Sivaraman et~al\mbox{.}}{2021}]%
        {sivaraman2021mining}
\bibfield{author}{\bibinfo{person}{Aishwarya Sivaraman}, \bibinfo{person}{Rui
  Abreu}, \bibinfo{person}{Andrew Scott}, \bibinfo{person}{Tobi Akomolede},
  {and} \bibinfo{person}{Satish Chandra}.} \bibinfo{year}{2021}\natexlab{}.
\newblock \showarticletitle{Mining Idioms in the Wild}.
\newblock \bibinfo{journal}{\emph{arXiv preprint arXiv:2107.06402}}
  (\bibinfo{year}{2021}).
\newblock


\bibitem[\protect\citeauthoryear{Slatkin}{Slatkin}{2019}]%
        {slatkin2019effective}
\bibfield{author}{\bibinfo{person}{Brett Slatkin}.}
  \bibinfo{year}{2019}\natexlab{}.
\newblock \bibinfo{booktitle}{\emph{Effective python: 90 specific ways to write
  better python}}.
\newblock \bibinfo{publisher}{Addison-Wesley Professional}.
\newblock


\bibitem[\protect\citeauthoryear{Tsantalis, Mansouri, Eshkevari, Mazinanian,
  and Dig}{Tsantalis et~al\mbox{.}}{2018}]%
        {tsantalis2018accurate}
\bibfield{author}{\bibinfo{person}{Nikolaos Tsantalis}, \bibinfo{person}{Matin
  Mansouri}, \bibinfo{person}{Laleh Eshkevari}, \bibinfo{person}{Davood
  Mazinanian}, {and} \bibinfo{person}{Danny Dig}.}
  \bibinfo{year}{2018}\natexlab{}.
\newblock \showarticletitle{Accurate and efficient refactoring detection in
  commit history}. In \bibinfo{booktitle}{\emph{2018 IEEE/ACM 40th
  International Conference on Software Engineering (ICSE)}}. IEEE,
  \bibinfo{pages}{483--494}.
\newblock


\bibitem[\protect\citeauthoryear{Tsantalis, Mazinanian, and Rostami}{Tsantalis
  et~al\mbox{.}}{2017}]%
        {tsantalis2017clone}
\bibfield{author}{\bibinfo{person}{Nikolaos Tsantalis}, \bibinfo{person}{Davood
  Mazinanian}, {and} \bibinfo{person}{Shahriar Rostami}.}
  \bibinfo{year}{2017}\natexlab{}.
\newblock \showarticletitle{Clone refactoring with lambda expressions}. In
  \bibinfo{booktitle}{\emph{2017 IEEE/ACM 39th International Conference on
  Software Engineering (ICSE)}}. IEEE, \bibinfo{pages}{60--70}.
\newblock


\bibitem[\protect\citeauthoryear{Van~Rossum et~al\mbox{.}}{Van~Rossum
  et~al\mbox{.}}{2007}]%
        {van2007python}
\bibfield{author}{\bibinfo{person}{Guido Van~Rossum} {et~al\mbox{.}}}
  \bibinfo{year}{2007}\natexlab{}.
\newblock \showarticletitle{Python Programming language.}. In
  \bibinfo{booktitle}{\emph{USENIX annual technical conference}},
  Vol.~\bibinfo{volume}{41}. \bibinfo{pages}{1--36}.
\newblock


\bibitem[\protect\citeauthoryear{Wang, Li, Liu, and Cai}{Wang
  et~al\mbox{.}}{2020}]%
        {wang2020exploring}
\bibfield{author}{\bibinfo{person}{Jiawei Wang}, \bibinfo{person}{Li Li},
  \bibinfo{person}{Kui Liu}, {and} \bibinfo{person}{Haipeng Cai}.}
  \bibinfo{year}{2020}\natexlab{}.
\newblock \showarticletitle{Exploring how deprecated python library apis are
  (not) handled}. In \bibinfo{booktitle}{\emph{Proceedings of the 28th acm
  joint meeting on european software engineering conference and symposium on
  the foundations of software engineering}}. \bibinfo{pages}{233--244}.
\newblock


\bibitem[\protect\citeauthoryear{Wei{\ss}gerber and Diehl}{Wei{\ss}gerber and
  Diehl}{2006}]%
        {weissgerber2006identifying}
\bibfield{author}{\bibinfo{person}{Peter Wei{\ss}gerber} {and}
  \bibinfo{person}{Stephan Diehl}.} \bibinfo{year}{2006}\natexlab{}.
\newblock \showarticletitle{Identifying refactorings from source-code changes}.
  In \bibinfo{booktitle}{\emph{21st IEEE/ACM international conference on
  automated software engineering (ASE'06)}}. IEEE, \bibinfo{pages}{231--240}.
\newblock


\bibitem[\protect\citeauthoryear{Xing and Stroulia}{Xing and Stroulia}{2006}]%
        {xing2006refactoring}
\bibfield{author}{\bibinfo{person}{Zhenchang Xing} {and} \bibinfo{person}{Eleni
  Stroulia}.} \bibinfo{year}{2006}\natexlab{}.
\newblock \showarticletitle{Refactoring detection based on umldiff change-facts
  queries}. In \bibinfo{booktitle}{\emph{2006 13th Working Conference on
  Reverse Engineering}}. IEEE, \bibinfo{pages}{263--274}.
\newblock


\bibitem[\protect\citeauthoryear{Zhang, Huang, Xia, Zou, Lo, and Xing}{Zhang
  et~al\mbox{.}}{2020}]%
        {zhang2020chatbot4qr}
\bibfield{author}{\bibinfo{person}{Neng Zhang}, \bibinfo{person}{Qiao Huang},
  \bibinfo{person}{Xin Xia}, \bibinfo{person}{Ying Zou}, \bibinfo{person}{David
  Lo}, {and} \bibinfo{person}{Zhenchang Xing}.}
  \bibinfo{year}{2020}\natexlab{}.
\newblock \showarticletitle{Chatbot4qr: Interactive query refinement for
  technical question retrieval}.
\newblock \bibinfo{journal}{\emph{IEEE Transactions on Software Engineering}}
  (\bibinfo{year}{2020}).
\newblock


\end{thebibliography}
